\documentclass[aps,prd,tightenlines,nofootinbib,reprint,superscriptaddress]{revtex4-1}

\usepackage{graphicx}
\usepackage{dcolumn}
\usepackage{bm}
\usepackage{amsmath}
\usepackage{amssymb}
\usepackage{multirow}

\begin{document}

\newcommand{\mevcc}{\!\mathrm{MeV}\!/c^2}
\newcommand{\mevc}{\!\mathrm{MeV}/\!c}
\newcommand{\mev}{\!\mathrm{MeV}}
\newcommand{\gevcc}{\!\mathrm{GeV}/\!c^2}
\newcommand{\gevc}{\!\mathrm{GeV}/\!c}
\newcommand{\gev}{\!\mathrm{GeV}}

\title{Precision Measurement of the Hadronic Contribution to the Muon Anomalous Magnetic Moment}

\author{T.~Xiao}
\author{S.~Dobbs}
\altaffiliation{Current affiliation: Florida State University, Tallahassee, Florida 32306, USA}
\author{A.~Tomaradze}
\author{Kamal~K.~Seth}

\affiliation{Northwestern University, Evanston, Illinois 60208, USA}

\author{G. Bonvicini}
\affiliation{Wayne State University, Detroit, Michigan 48202, USA}

\date{December 11, 2017}

\begin{abstract} 
We report on a precision measurement of the cross section for the reaction $e^+e^-\to\pi^+\pi^-$ in the mass range $0.30<M_{\pi\pi}<1.00$~GeV with the initial state radiation (ISR) method, using 817~pb$^{-1}$ of data at $e^+e^-$ center-of-mass energies near 3.77~GeV and 586~pb$^{-1}$ of data at $e^+e^-$ center-of-mass energies near 4.17~GeV, collected with the CLEO-c detector at the CESR $e^+e^-$ collider at Cornell University.
The integrated cross sections in the range $0.30<M_{\pi\pi}<1.00$~GeV for the process $e^+e^-\to\pi^+\pi^-$ are determined with a statistical uncertainty of $0.7\%$ and a systematic uncertainty of $1.5\%$. 
The leading-order hadronic contribution to the muon anomalous magnetic moment calculated using these measured $e^+e^-\to\pi^+\pi^-$ cross sections in the range $M_{\pi\pi}=0.30$ to 1.00~GeV is calculated to be $(500.4\pm3.6~(\mathrm{stat})\pm~7.5(\mathrm{syst}))\times10^{-10}$.
\end{abstract}

\maketitle

\section{Introduction} \label{intro}

Magnetic moments of leptons offer among the most promising opportunities to provide critical tests of the standard model (SM) of particle physics because both the experimental measurements and the theoretical predictions can be made with very high level of precision. The magnetic moment of the electron has been measured by the Harvard University group of Gabrielse~\cite{hanneke}, and expressed in terms of the ``anomaly'', $a_e^{\mathrm{exp}}\equiv(g_e-2)/2$, which is $a_e^{\mathrm{exp}}=1,159,652,180.73~(28)\times10^{-12}$.
The theoretical prediction based on 10th-order QED calculation is $a_e^{\mathrm{QED}} = 1,159,652,179.936~(764)\times10^{-12}$~\cite{kinoshita3}.
Because electrons do not decay, the non-QED corrections to $a_e$ are extremely small, $a_e^{\mathrm{had,weak}}=1.707~(16)\times10^{-12}$~\cite{vp,ll,weak1,weak2,weak3,weak4}, bringing the Standard Model prediction to $a_e^{\mathrm{SM}} = 1,159,652,181.643~(764)\times10^{-12}$,
and the difference, $\Delta a_e\equiv a_e^{\mathrm{SM}}-a_e^{\mathrm{exp}}=0.91~(82)\times10^{-12}$.
This superb level of agreement is rightly considered the crowning achievement for both QED and the SM. An even higher level of achievement can be realized by considering the magnetic moment of the next heavier lepton, the muon, because lepton universality is considered to be well established, and sensitivity to `beyond the standard model (BSM)' effects is expected to increase with lepton mass. However, the non-QED contributions to $a_\mu$ are much larger than for $a_e$, and some of them can not be reliably calculated theoretically; they need to be experimentally measured with precision. In this paper we report on a precision measurement of the largest such non-QED contribution to the anomalous magnetic moment of the muon, $a_\mu^{\mathrm{SM}}$, the contribution $a_\mu^{\mathrm{had,LO}}$ due to hadronic contributions at low energy.

The QED prediction for the muon magnetic moment (in units of $10^{-11}$) calculated to the 10th-order~\cite{qedlevel} is $a_{\mu}^{\mathrm{QED}}=116,584,718.846~(36)$.
Unlike the electron, the muon magnetic moment receives substantial contributions from higher-order contributions involving virtual weak and hadronic particles. The largest of these is the lowest-order hadronic contribution, $a_\mu^{\mathrm{had,LO}}$, which can not be calculated reliably by theory and needs to be measured experimentally, as we do in the present paper. Other contributions are smaller, and have been theoretically calculated. These are:
electroweak contributions, $a_\mu^{\mathrm{weak}}=153.6~(10)$~\cite{ewlevel}, higher order hadronic contributions, $a_\mu^{\mathrm{had,HO}}=-98.4~(7)$~\cite{holevel}, and ``light-by-light'' contributions, $a_\mu^{\mathrm{had,LBL}}=116~(39)$~\cite{lbllevel}, so that $a_\mu^{\mathrm{SM}}=116,584,890~(39)+a_\mu^{\mathrm{had,LO}}$.
In this paper we report the results of our measurements of $a_\mu^{\mathrm{had,LO}}$.

The anomalous magnetic moment contribution, $a_\mu^{\mathrm{had,LO}}$ is related to the measured Born-level cross section, $\sigma_0(e^+e^-\to\mathrm{hadrons})$ via the dispersion relation~\cite{dispersion}
\begin{equation}\label{eq:eq165}
a_\mu^{\mathrm{had,LO}}=\frac{\alpha^2(0)}{3\pi^2}\int_{4m_\pi^2}^{\infty}ds\frac{K(s)}{s}\frac{\sigma_0(s)}{\sigma(\mathrm{pt})},
\end{equation}
where $s$ is the center of mass energy, $K(s)$ is the QED kernel~\cite{kernel},
\begin{equation}\begin{split}
K(s)=&x^2(1-\frac{x^2}{2})+(1+x)^2(1+\frac{1}{x^2})(\ln(1+x)-x+\frac{x^2}{2}) \\
          &+\frac{(1+x)}{1-x}x^2\ln x,
\end{split}
\end{equation}
$x=(1-\beta_\mu)/(1+\beta_\mu)$, $\beta_\mu=(1-4m_\mu^2/s)^{1/2}$, and
$\sigma(\mathrm{pt}) = 4\pi\alpha^2/3s$. 
In principle, to calculate this integral, the hadronic cross sections should be measured at all center-of-mass energies ($\sqrt{s}$), and for all possible hadronic decay final states, beginning with the lowest-mass final state of two pions. 
However, cross sections generally decrease with increasing $s$, increasing multiplicity and hadron masses, and the QED kernel also decreases monotonically with $s$. As a result, most of the contribution to $a_\mu^{\mathrm{had,LO}}$ comes from the hadronic cross section at small center-of-mass energy, with $\sim 91\%$ of it coming from $\sqrt{s}<1.8$~GeV, and $73\%$ from the $\pi^+\pi^-$ final state~\cite{hvptools}. In view of this, most experimental measurements of $a_\mu^{\mathrm{had,LO}}$ have been focused on $\sqrt{s}\leq 1$~GeV, and especially on $\pi^+\pi^-$ decays.
These consist of measurements by the Novosibirsk Collaborations CMD-2~\cite{cmd2} and SND~\cite{snd} by varying primary $e^+e^-$ collision energies in the region $0.36<\sqrt{s}<1.4$~GeV, 
and by using the initial state radiation (ISR) method to obtain varying effective center-of-mass energies by BaBar~\cite{babar2} (26 final states, $\sqrt{s}<5$~GeV), by KLOE~\cite{kloe} ($\pi^+\pi^-,~\sqrt{s}\lesssim1.0$~GeV), and most recently by BES-III~\cite{besiii} ($\pi^+\pi^-$, in the limited range $0.6<\sqrt{s}<0.9$~GeV).

\begin{figure}
\begin{center}
\includegraphics[width=3.5in]{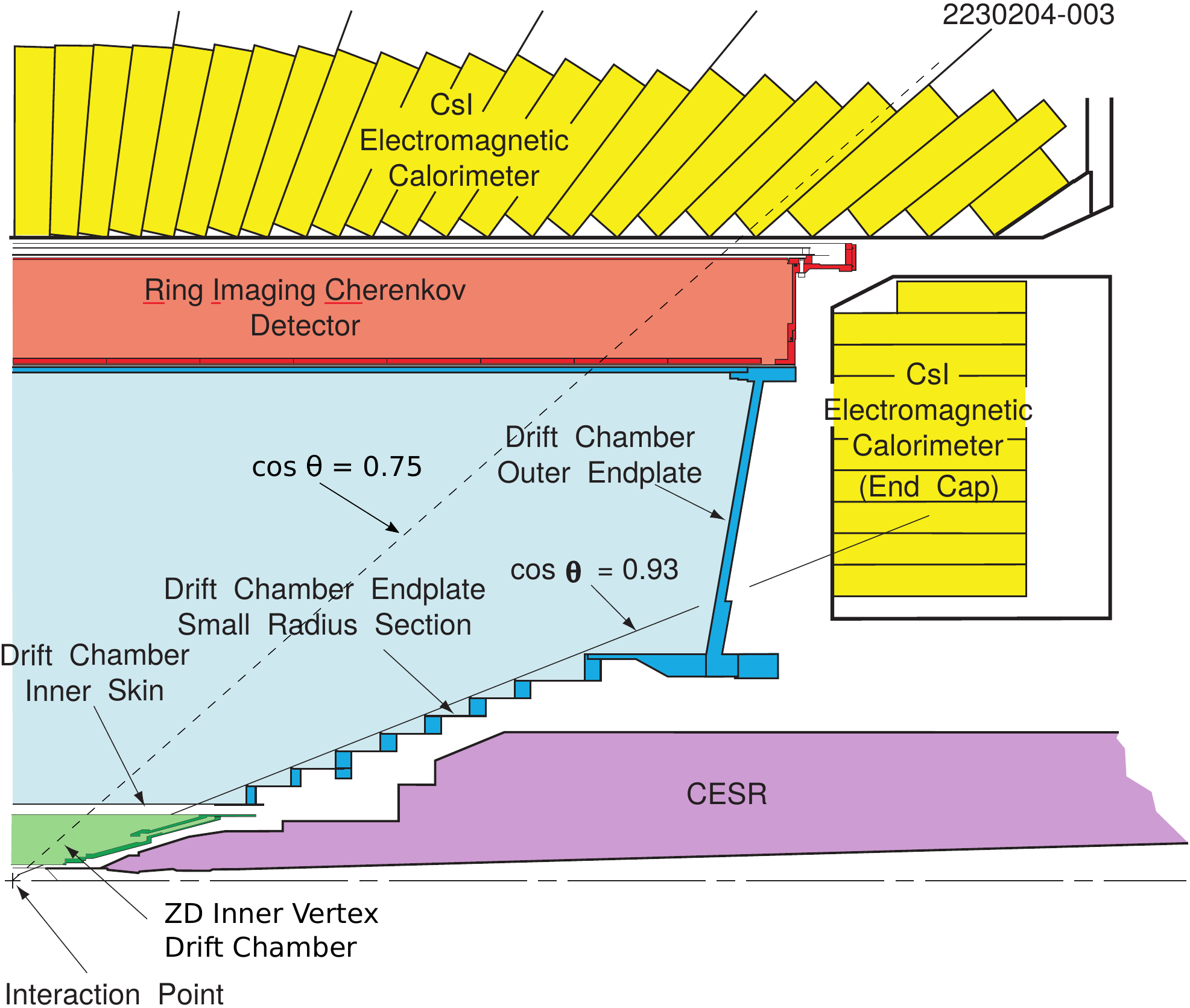}
\end{center}
\caption{Schematic representation of important components of the CLEO-c detector.}
\label{fig:cleodet}
\end{figure}

In this paper, we also use the ISR method to measure $\sigma(e^+e^-\to\pi^+\pi^-)$ in the region $0.30\leq\sqrt{s}<1.00$~GeV.
We use data for $e^+e^-$ annihilations taken at the CESR collider, 817~pb$^{-1}$ at $\sqrt{s}=3.77$~GeV, and 586~pb$^{-1}$ at $\sqrt{s}=4.17$~GeV. The resulting particles were detected in the CLEO-c detector which has been described in detail elsewhere~\cite{cleodetector}. The detector has a cylindrically symmetric configuration, and it provides $93\%$ coverage of solid angle for charged and neutral particle identification. The detector components important for the present measurements are the vertex drift chamber, the main drift chamber, the CsI(Tl) crystal calorimeter (CC), and the Ring Imaging Cherenkov detector (RICH), which are illustrated in Fig~\ref{fig:cleodet}.
Simulated event samples of the signal ISR process $e^+e^-\to \pi^+\pi^-\gamma_{\mathrm{ISR}}$, the main background ISR process $e^+e^-\to \mu^+\mu^-\gamma_{\mathrm{ISR}}$, and other background multihadronic ISR processesare generated using the Monte Carlo (MC) event generator Phokhara~\cite{phokhara09}.
Background ISR processes $e^+e^-\to \psi(1S,2S)\gamma_{\mathrm{ISR}}$ are simulated with the MC event generator EvtGen~\cite{evtgen}, and background events from $e^+e^-\to q\bar{q}~(q=u,d,s)$ are generated using the LUNDCHARM model~\cite{lundcharm} implemented in EvtGen.
In each case, the events are then passed through a GEANT-based~\cite{geant} detector simulation of the CLEO-c detector, where the simulated detector response is reconstructed in the same way as for the actual data.

\section{Event Selections} \label{selection}

In order to reconstruct $e^+e^- \to \pi^+\pi^- \gamma_\text{ISR}$ events, we select events with only two reconstructed oppositely-charged tracks and at least one reconstructed calorimeter shower. 
The highest energy shower is assumed to be due to the ISR photon. This shower is required to have an energy larger than 0.5~GeV, and have a transverse energy distribution consistent with that of an electromagnetic shower.
In this analysis, pions are primarily identified by their energy deposition in the crystal calorimeter (CC). 
For charged particles we select events with polar angle $|\cos\theta|<0.75$ for which the CC has the most uniform response and is best understood.
The efficiency of pions to pass the CC criteria is calculated using a sample of inclusive pions from $\sim48~\mathrm{pb}^{-1}$ of data taken at the $\psi(2S)$ resonance. To ensure high precision in determining efficiency, we require the momenta of both tracks to be $p<1.6$~GeV.
In addition, for events in the $M_{\pi\pi}$ region below 0.5~GeV we require the transverse momentum of the charged particle tracks to be $p_T>0.2$~GeV in order to reduce the contribution of poorly reconstructed tracks in this region.

\subsection{Particle Identification} \label{pid}

For particle identification the observables are: $p$, the momentum measured in the drift chambers; $E_{\mathrm{CC}}$, the energy deposited in the central calorimeter; $dE/dx$, the ionization energy deposited in the main drift chamber, and for $p>0.5$~GeV the Cherenkov photons detected in the RICH  detector.

The main backgrounds to the  $\pi^+\pi^- \gamma_\text{ISR}$ final state are similar reactions, with the charged pions replaced by charged kaons, electrons, or muons.  

\textbf{$\bm{\pi/K}$ separation:}
We distinguish between charged pions and kaons using only the energy loss in the drift chamber ($dE/dx$) at low momentum ($p<0.5$~GeV), or a combined likelihood variable using $dE/dx$ and $LL^{\mathrm{RICH}}$ at higher momentum ($p>0.5$~GeV), where $LL^{\mathrm{RICH}}$ is the log-likelihood that a particle corresponds with a given hypothesis based on Cherenkov photons detected in the RICH detector.  We reject any event with a track that is found to be more ``kaon-like'' than ``pion-like'' using these criteria. 
Specifically, for low momentum tracks, we separate kaon candidates from pion candidates by requiring
$|\sigma^{dE/dx}_K|<3$ and $|\sigma^{dE/dx}_K|<|\sigma^{dE/dx}_\pi|$.
For high momentum tracks, we use the combined likelihood variable
\begin{equation}
\Delta L_{K,\pi}=(LL^{\mathrm{RICH}}_K-LL^{\mathrm{RICH}}_\pi)+((\sigma^{dE/dx}_K)^2-(\sigma^{dE/dx}_\pi)^2).
\end{equation}
We reject events that have a track with $\Delta L_{K,\pi}<0$.

\textbf{$\bm{\pi/e}$ separation:}
The ratio $E_{\mathrm{CC}}/p$ for electrons is $\sim1.0$, while for pions it is generally much smaller because pions do not electromagnetically shower. 
Therefore we can efficiently reject electrons by removing events with a charged track which has corresponding $E_{\mathrm{CC}}/p>0.8$.

\begin{figure}[!tb]
\begin{center}
\includegraphics[width=3.5in]{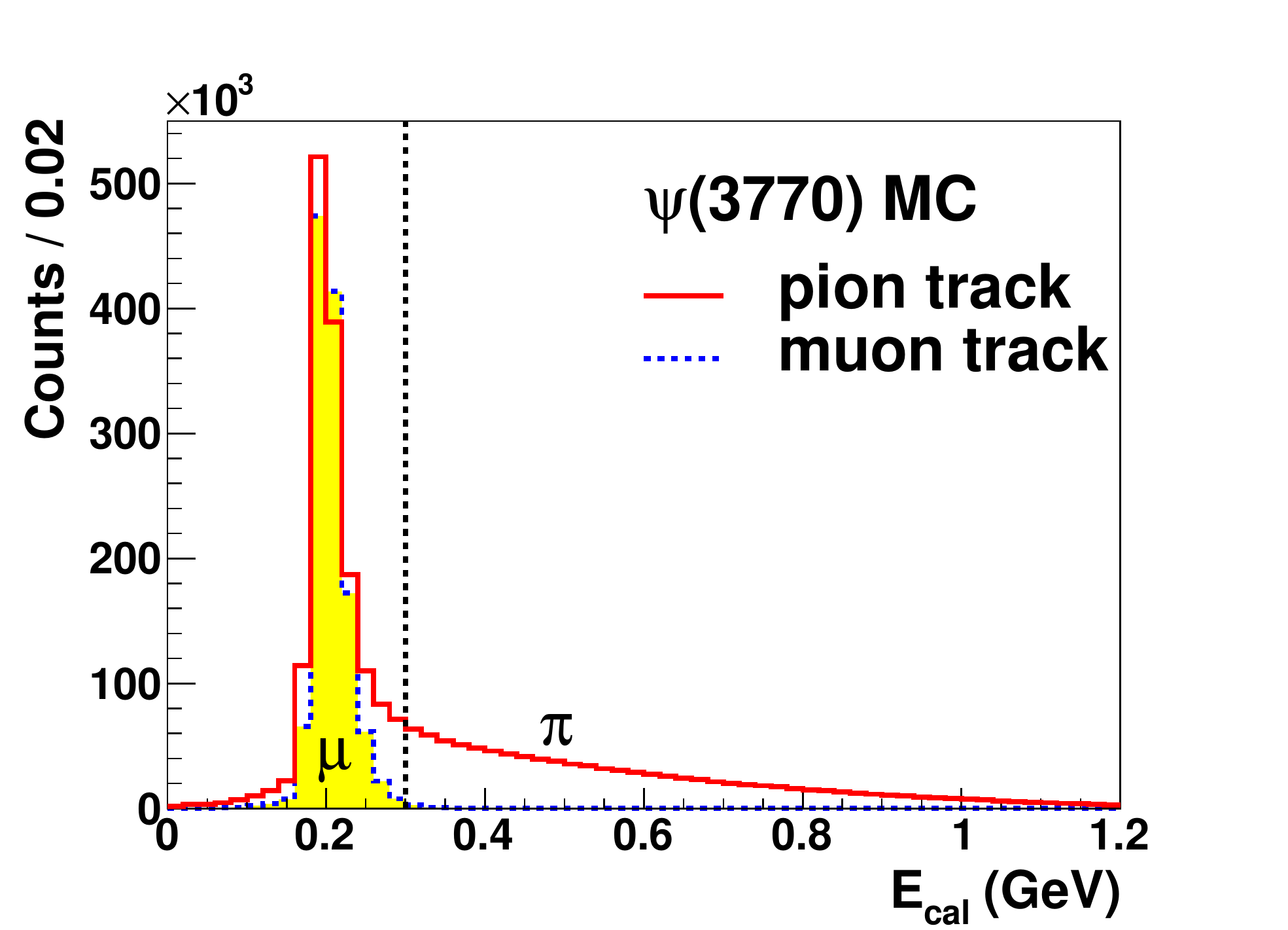}
\includegraphics[width=3.5in]{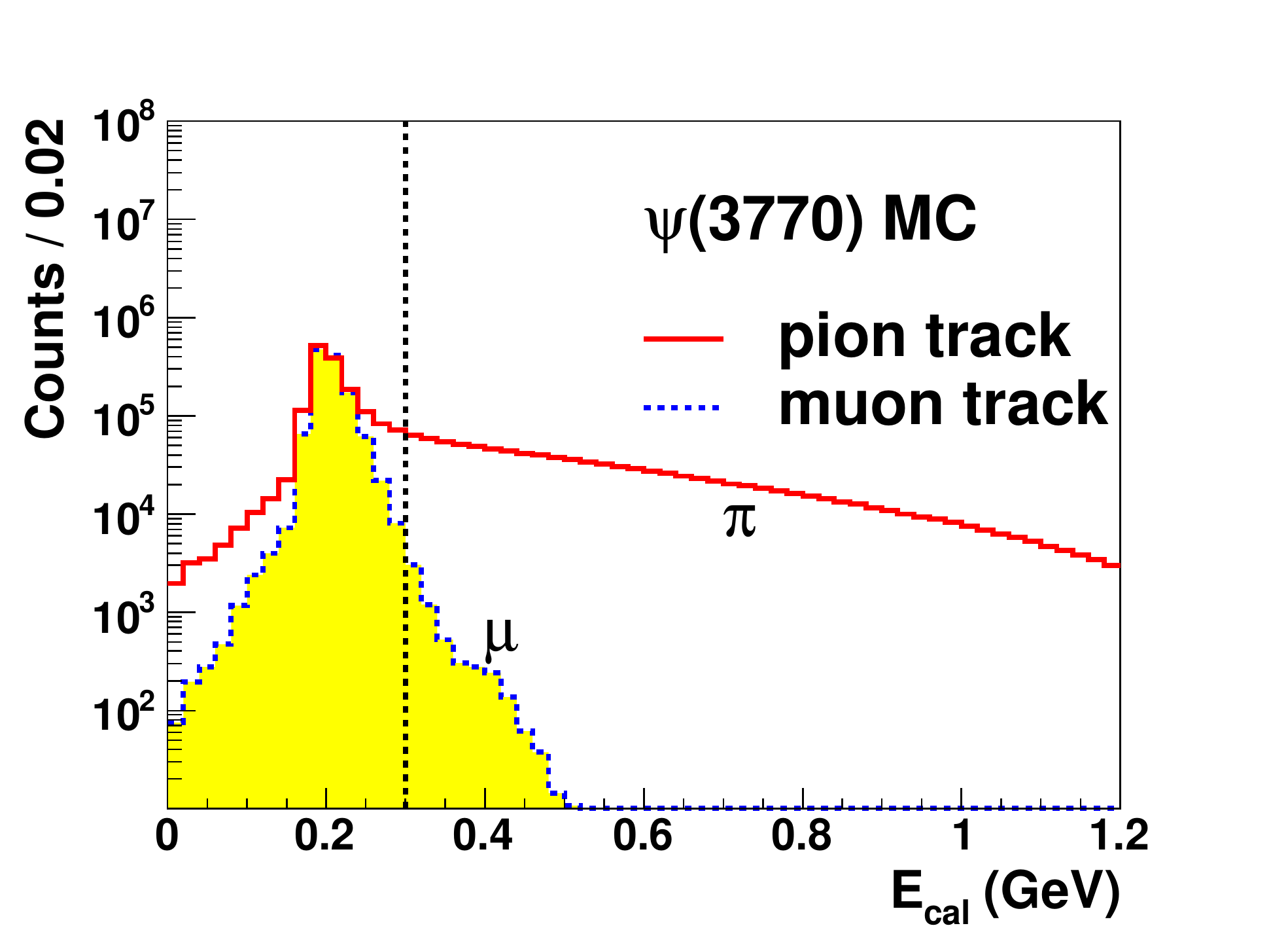}
\end{center}
\caption[The shower energy deposited in the CC associated with a pion track or a muon track.]
{Monte Carlo simulation of energy deposit in the CC by muon and pion tracks. Top: linear plot; bottom: log plot.}
\label{fig:matched}
\end{figure}

\textbf{$\bm{\pi/\mu}$ separation:}
$\pi/\mu$ separation by $dE/dx$ in the drift chamber is challenging because the pions and muons have nearly equal masses and therefore deposit similar amounts of electromagnetic energy. 
However, because pions do also interact by the strong interaction, they deposit additional energy in the CC, which shows up as a long  tail in the energy deposited. This is illustrated in Fig.~\ref{fig:matched}.
A good rejection of muons is achieved by rejecting events in which both charged particles have $E_\mathrm{CC}<0.3$ GeV. This requirement rejects $\sim98\%$ of $\mu^+\mu^-$ events, but unfortunately it also results in the loss of nearly $50\%$ of $\pi^+\pi^-$ events.
For some events at low $M_{\pi\pi}$, the two tracks produce overlapping energy deposits in the CC, which add up to $E_\text{CC}>0.3$~GeV.
These events amount to about $1\%$ of our total event sample, and $5\%$ of the events in the $0.4<M_{\pi\pi}<0.55$~GeV region.
We reject these events because $\mu^+\mu^-\gamma_{\mathrm{ISR}}$ events would be mis-identified as $\pi^+\pi^-\gamma_{\mathrm{ISR}}$ events with the overlapping energy deposits of two muons appearing to be energy loss due to a single pion. 

\subsection{$\pi^0$ Rejection}

\begin{figure}[!tb]
\begin{center}
\includegraphics[width=3.5in]{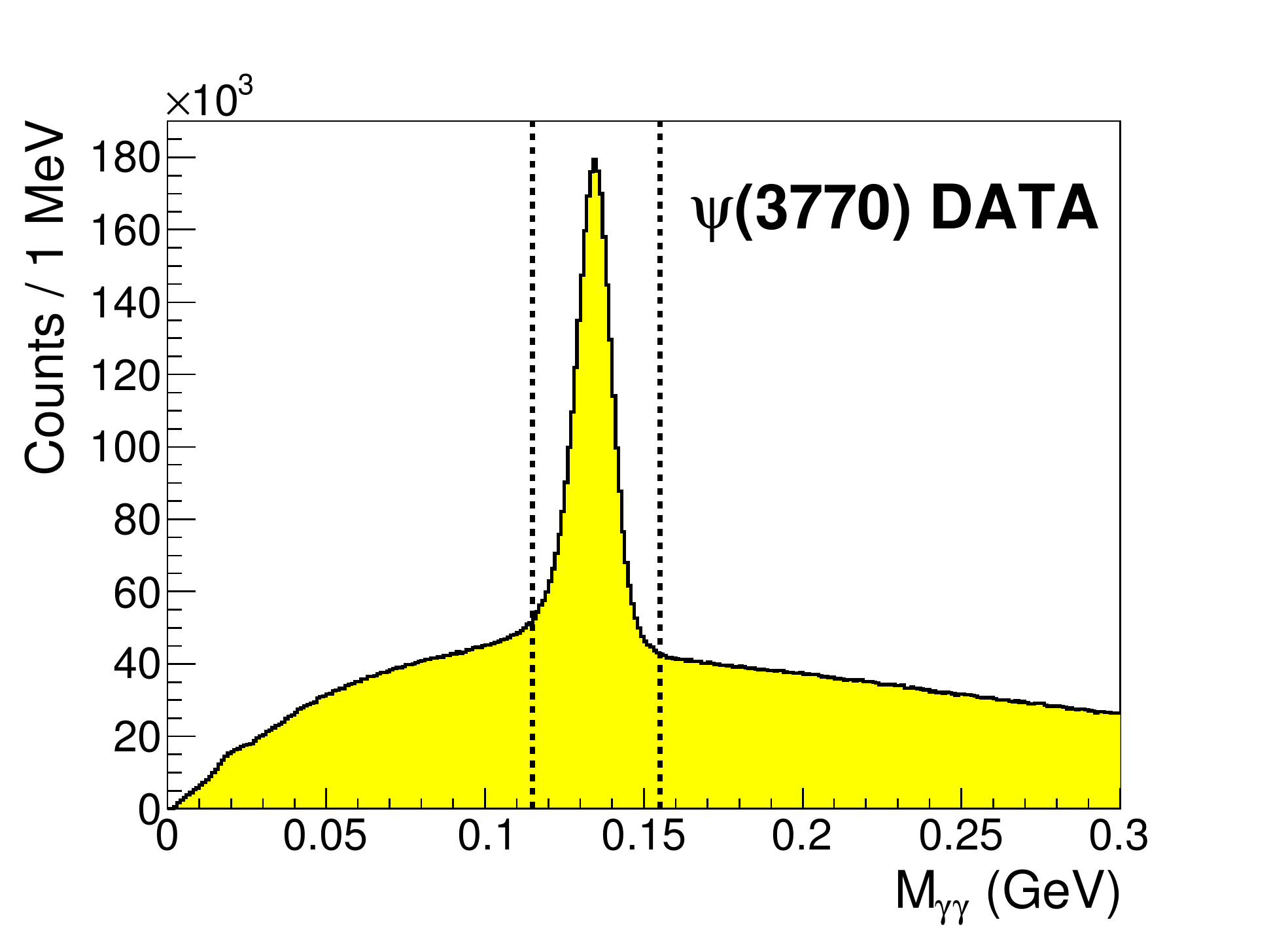}
\end{center}
\caption[Two-photon invariant mass distribution from data.]
{Two-photon invariant mass distribution from data, illustrating the $\pi^0$ peak. The $\pi^0$ rejection cut of $M_{\gamma\gamma}=M_{\pi^0}\pm20$~MeV is indicated by the vertical lines.}
\label{fig:pi0_cut}
\end{figure}

To reduce the contribution from hadronic decays containing photons from $\pi^0\to\gamma\gamma$ decays, where the $\pi^0$ decays asymmetrically and one of the photons may be confused with the ISR photon, events are rejected if a pair of photons in the event forms a $\pi^0$ candidate with mass within $20~\mathrm{MeV}$ of the nominal $\pi^0$ mass of 135~MeV. The $M_{\gamma\gamma}$ distribution and the $\pi^0$ rejection cut are illustrated in Fig.~\ref{fig:pi0_cut}.

\subsection{Kinematic Fit}

\begin{figure}[!tb]
\begin{center}
\includegraphics[width=2.in]{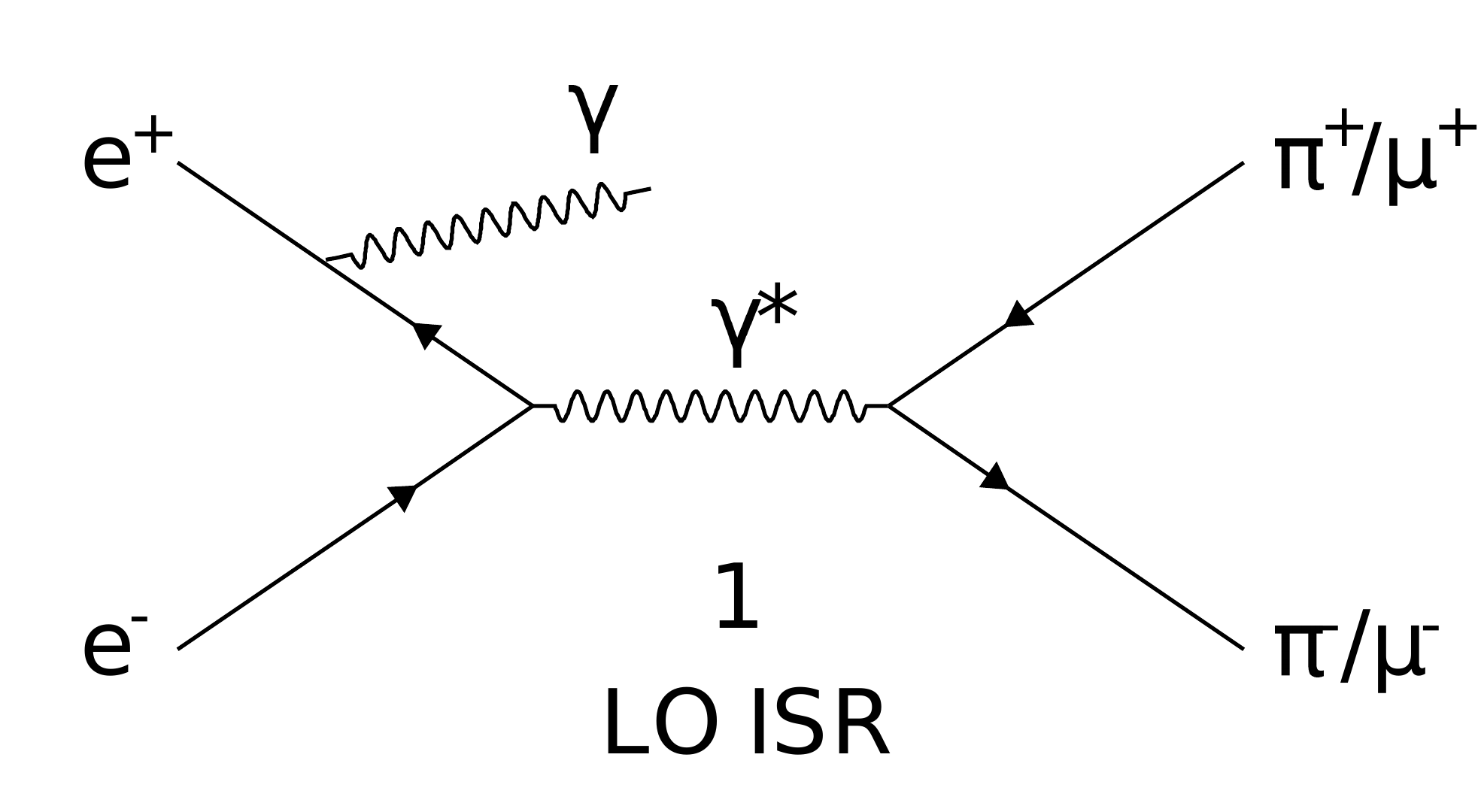}
\includegraphics[width=2.in]{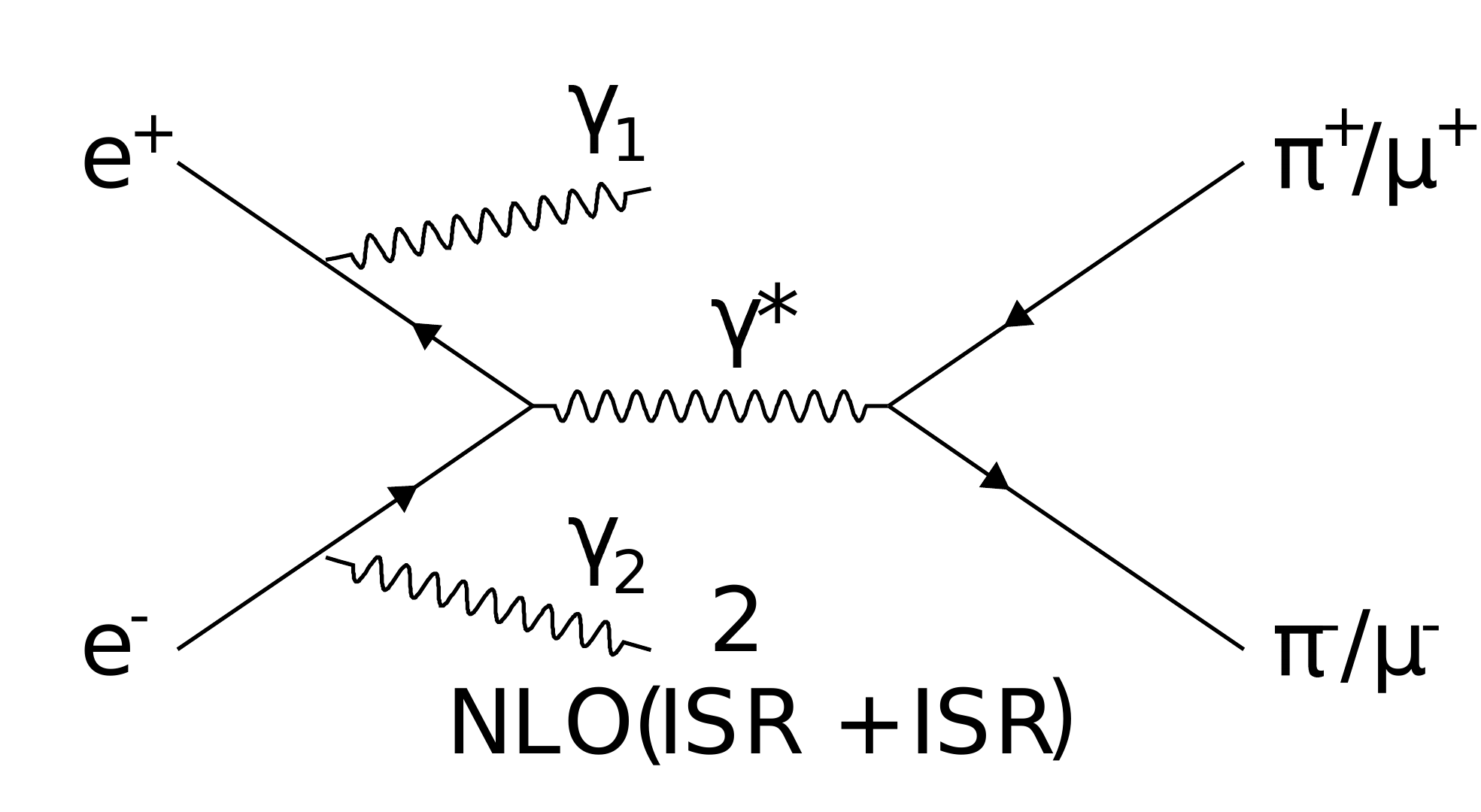}
\includegraphics[width=2.in]{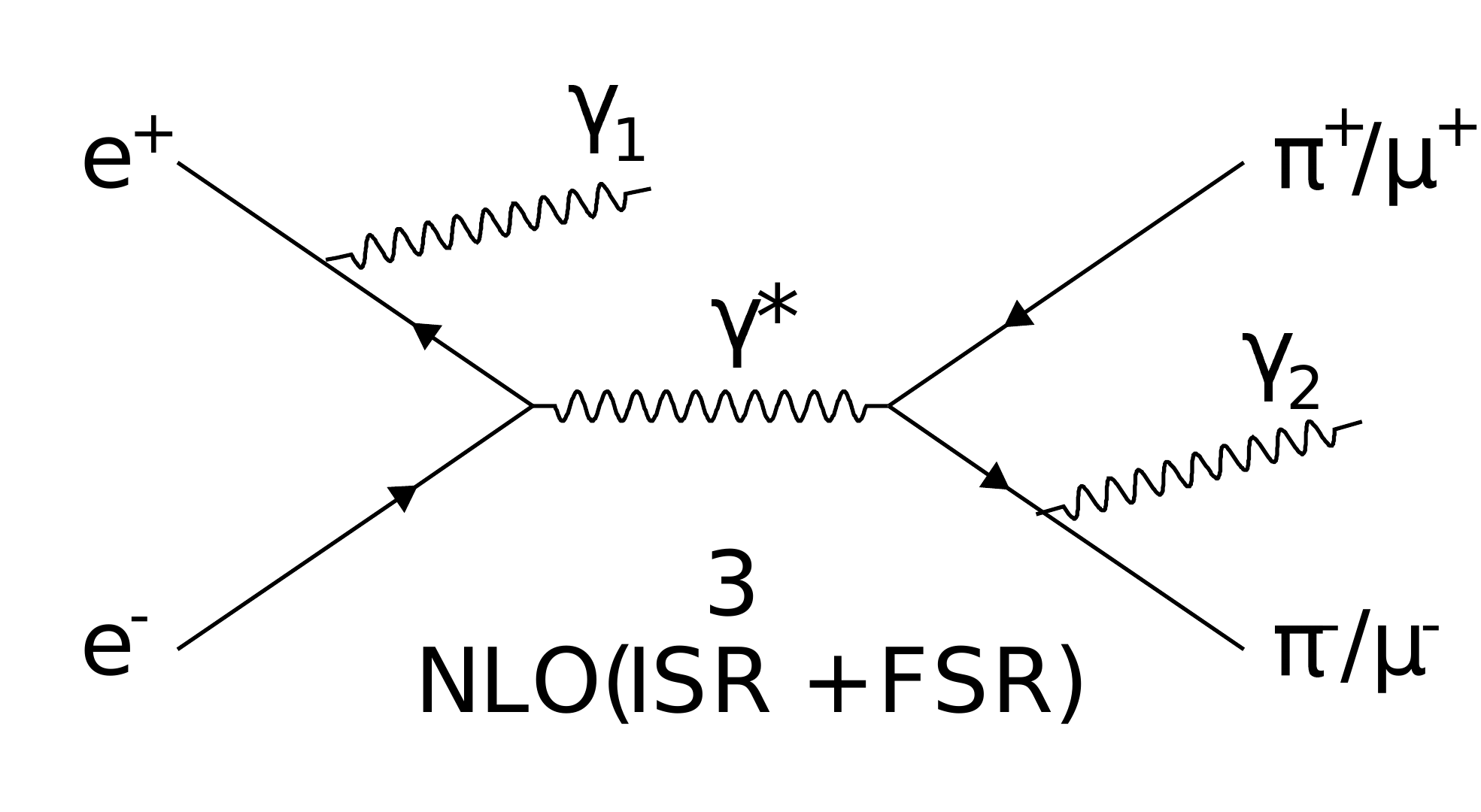}
\end{center}
\caption[The Feynman diagrams for the ISR processes relevant to this study.]
{The Feynman diagrams for the ISR processes relevant to this study: 1. LO ISR, 2. NLO with additional ISR, 3. NLO with additional FSR.}
\label{fig:addphoton}
\end{figure}

In order to measure $e^+e^- \to \pi^+\pi^-\gamma_\text{ISR}$ cross sections down to a percent-level accuracy, it is necessary to include the next-to-leading order (NLO) correction.
So we keep events which have a radiative photon in addition to the primary ISR photon.
The additional photon can be either an ISR or a final-state radiation (FSR) photon.
Fig.~\ref{fig:addphoton} shows the Feynman diagrams of LO and NLO ISR processes which are selected in this study.

\begin{figure}[!tb]
\begin{center}
\includegraphics[width=3.2in]{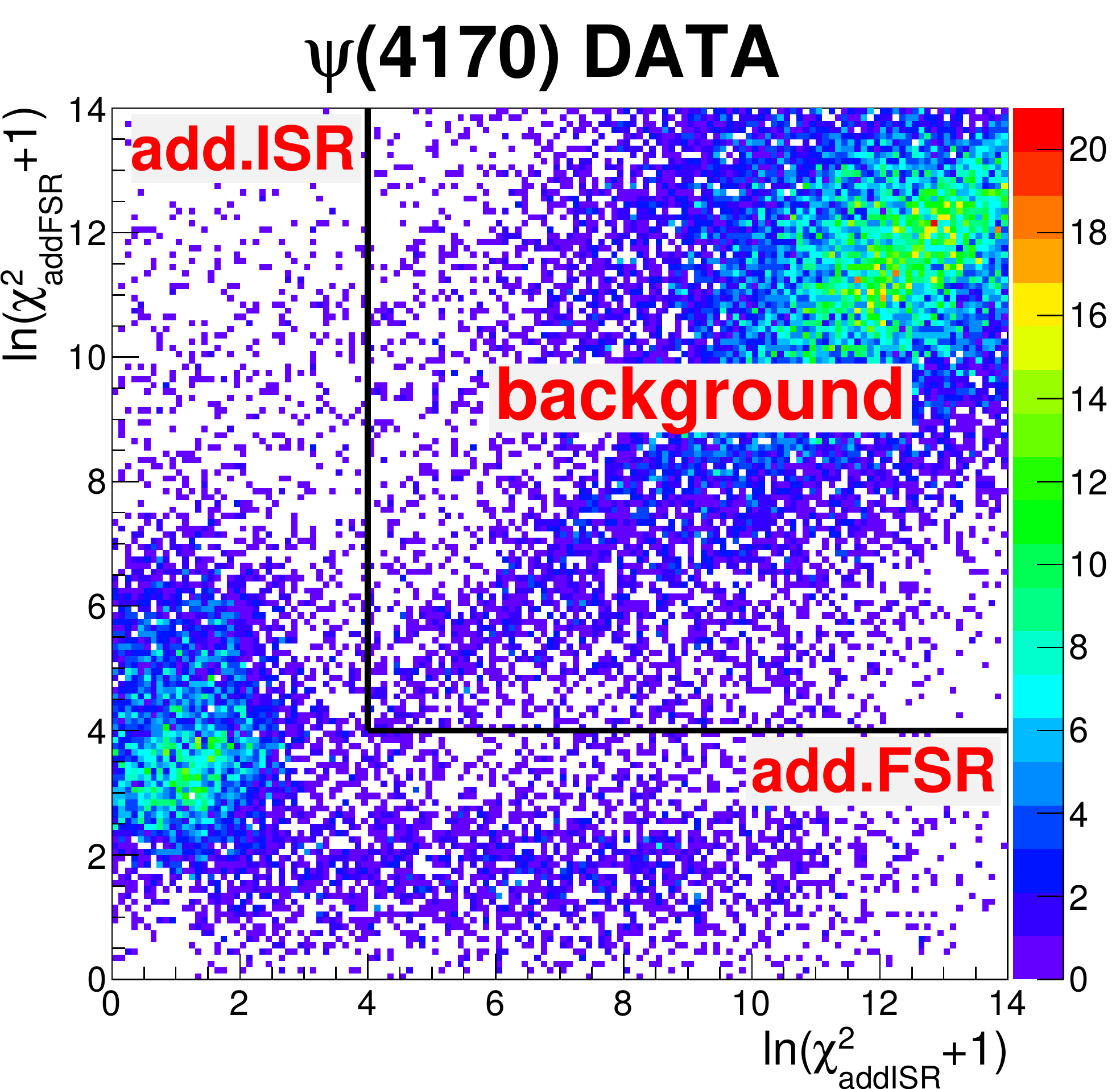}
\includegraphics[width=3.2in]{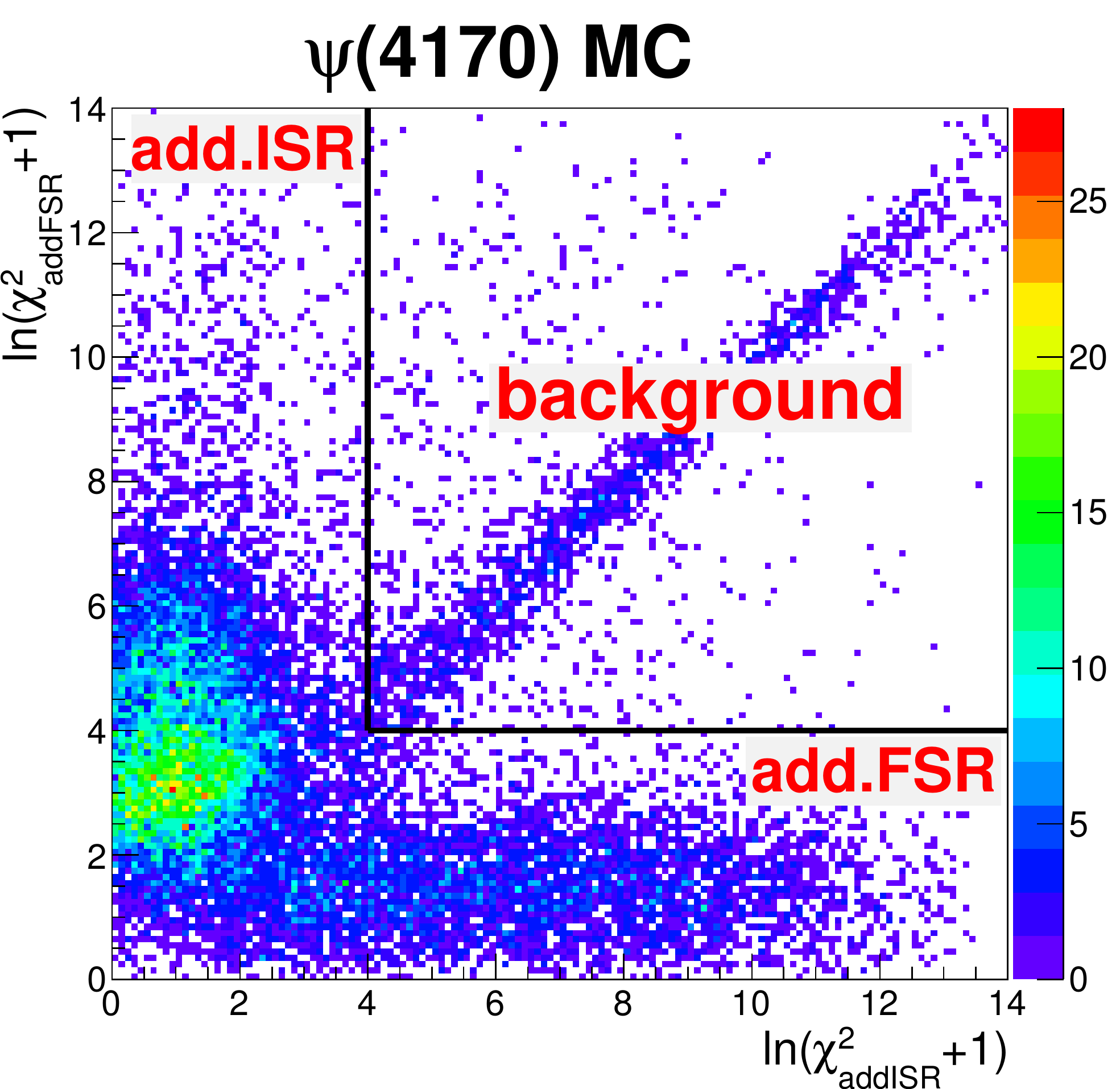}

\end{center}
\caption
{The 2D distributions for $\ln(\chi^2+1)$ for $\pi^+\pi^-\gamma_\mathrm{ISR}$ plus $\mu^+\mu^-\gamma_\mathrm{ISR}$ for $\psi(4170)$ data and signal MC. The vertical and horizontal lines denote the boundaries for the $\chi^2$ selection; events to the left of the vertical line, and below the horizontal line are accepted as signal events. Background events in the box region are rejected. A similar plot for the $\psi(3770)$ data is not shown.}
\label{fig:2Dfit_4170}
\end{figure}

To reconstruct the full $ \pi^+\pi^-\gamma_\text{ISR}$ event including a possible additional radiative photon from ISR or FSR, following BaBar~\cite{babar2} we perform two types of kinematic fits constraining the final state particles to have the $e^+e^-$ collision center-of-mass four-momentum. These two kinematic fits are:
\begin{itemize}
\item Fit $\#1$ (additional FSR fit): 
If the additional photon is detected in the electromagnetic calorimeter, we call it an ``additional FSR'' fit, although the extra photon could be generated either from FSR or from ISR at large enough angle to be detectable in the CC.
We set the energy threshold for the additional photon at 30~MeV.
In case of several extra detected photons, we perform the kinematic fit using each photon in turn and the fit with the smallest $\chi^2_\mathrm{add.~FSR}$ is retained. In practice we plot $\chi^2_1 \equiv \ln(\chi^2_\mathrm{add.~FSR}+1)$ so that we can see the long tail.
\item Fit $\#2$ (additional ISR fit): 
This kinematic fit is done assuming an additional photon which is not detected in the electromagnetic calorimeter.
The events which are fit using this method include both those in which only one photon is generated in the final state, and those in which an additional photon is an ISR photon produced at an angle too small to the beams to be detected in the CC.
Any other additional photon candidate measured in the electromagnetic calorimeter is ignored in this fit. 
Single photon events fit by this procedure will yield a good $\chi^2_\mathrm{add.~ISR}$ with an additional undetected photon energy near zero. Similar to $\chi^2_1$, we use $\chi^2_2 \equiv \ln(\chi^2_\mathrm{add.~ISR}+1)$.
\end{itemize}
The momenta of the charged particle tracks and the direction of the main ISR photon are used in both fits.
In the case that the main ISR photon is the only photon detected in the event, we perform only the additional ISR fit to the event, and reject events with $\ln(\chi^2+1)>6$.
Fig.~\ref{fig:2Dfit_4170} shows two dimensional plots of $\chi^2_1$ versus $\chi^2_2$ for the $\psi(4170)$ data and its MC simulation.
These plots show large enhancements when both $\chi^2_1$ and $\chi^2_2$ are small.
There are also enhancements parallel to both ISR and FSR axes.
These indicate events with additional radiation, and are included as signals.
The events in the large boxed regions include multihadron events which are not simulated by MC, and events along the diagonals.
These are considered as being due to backgrounds and are rejected.
In the following studies and in final distributions, the $\pi^+\pi^-$ invariant mass is obtained using the fitted parameters of the two charged particles from the additional FSR fit if $\chi^2_1<\chi^2_2$, and from the additional ISR fit if $\chi^2_2<\chi^2_1$.

\section{Efficiency Studies} \label{efficiency}

In order to precisely measure the $e^+e^- \to \pi^+\pi^-$ cross sections, the event reconstruction efficiency is considered in detail.  
We determine the most important parts of the efficiency, $ \epsilon_\text{E/p}$ for $\pi/e$ discrimination and $ \epsilon_\text{CC}$ for $\pi/\mu$ discrimination from our  $\psi(2S)$ data,
while the efficiency due to acceptance and all other selections,  $ \epsilon_\text{MC}$, is determined by MC simulations together with small correction factors $C_\mathrm{trig}$, $C_\mathrm{track}$, $C_{\gamma}$, and $C_{K}$ (all $\gtrsim 0.99$). 
The overall efficiency $\epsilon$ is determined as
\begin{equation} \label{eq:eff}
\epsilon = \epsilon_\text{E/p}  \times \epsilon_\text{CC}  \times \epsilon_\text{MC}  \times C_\mathrm{trig} \times C_\mathrm{track} \times C_{\gamma} \times C_{K}, 
\end{equation}
where $ \epsilon_\text{E/p}$ and $\epsilon_\text{CC}$ are the efficiencies for $\pi/e$ and $\pi/\mu$ separations determined by using data samples from $\psi(2S)$ inclusive decays as described in Sec.~\ref{eff}, 
$\epsilon_\text{MC}$ is the MC-determined efficiency for all criteria except the $\pi/e$ and $\pi/\mu$ separation criteria,
and the factors $C_\mathrm{trig}$, $C_\mathrm{track}$, $C_{\gamma}$, and $C_{K}$ are small correction factors to take into account of differences between the data and MC samples, as described below in Sec.~\ref{trigger},~\ref{tr},~\ref{isr},~and \ref{kaon}.

We have carefully studied the differences between data and MC simulation for various event selection criteria using a sample of $\pi^+\pi^-\gamma_{\mathrm{ISR}}$ and $\mu^+\mu^-\gamma_{\mathrm{ISR}}$ events from both $\psi(3770)$ and $\psi(4170)$ data sets. All MC efficiencies are compared to the efficiencies obtained from the data sample, and small corrections are applied to the efficiencies accordingly.  Generally, the corrections to the MC-determined efficiency are found to not have any significant dependence on $M_{\pi\pi}$ and to be consistent between both $\psi(3770)$ and $\psi(4170)$ data sets.

\subsection{Trigger Efficiency} \label{trigger}

There are 16 Level 1 (L1) hardware trigger lines in the CLEO-c trigger system. Of these, our $\pi^+\pi^-\gamma_\text{ISR}$ signal events are primarily selected by three different trigger lines.
The definitions of these trigger lines contain partially overlapping criteria, and the efficiency for signal events in MC simulations to pass any one these three trigger lines is $>99\%$.
To estimate the overall efficiency of the hardware trigger, we select events in data and signal MC samples which pass one of these triggers, and then look at the efficiency of these events to pass one of the other two triggers. 
The results are found to be consistent between data and MC within statistical uncertainties, and we apply no correction to the efficiency from this source.

\subsection{Track Reconstruction Efficiency} \label{tr}

We calculate the efficiency for charged particle tracks to be reconstructed using a dedicated sample of events. We select one reconstructed track and a photon (assumed to be the ISR photon), and apply a 1C kinematic fit assuming that there is a missing particle with the mass of the charged pion. A sample of events with 1 to 4 tracks and $\chi^2_{\mathrm{1C~fit}}<10$ is selected to estimate the tracking efficiency. The fraction of the predicted tracks that are actually reconstructed in the tracking system, with a charge opposite to that of the primary reconstructed track, yields one track reconstruction efficiency.
The predicted track is required to lie within the tracking acceptance, but because of subsequent decays or secondary interactions its momentum and angles do not have to match the expected values. 
The average one-track reconstruction efficiency correction of the two data sets is $0.993(4)$.
Since there are two tracks in our event selections, we apply $C_\mathrm{track}=[0.993(4)]^2=0.986(6)$ to the $\pi^+\pi^-$ cross section calculated in this analysis. 

\subsection{ISR Photon Reconstruction Efficiency} \label{isr}

We calculate the ISR photon reconstruction efficiency using a method similar to that described above for the track reconstruction efficiency. Instead of one reconstructed track and the ISR photon, we select two oppositely-charged tracks and apply a 1C kinematic fit assuming that there is a missing particle of zero mass, i.e., a photon. A sample of events with 2 tracks and any number of showers with $\chi^2_{\mathrm{1C~fit}}<5$ is selected to estimate the ISR photon reconstruction efficiency.
The fraction of the predicted photons that are actually reconstructed in the CC and identified as a photon, yields the ISR photon reconstruction efficiency.
The predicted photon is required to lie within the shower acceptance, but its momentum and angles do not have to match the expected values.
The average values of $C_{\gamma_{\mathrm{ISR}}}=0.992(6)$ are applied to the calculation of the $\pi^+\pi^-$ cross section. 

\subsection{Charged Pion Identification Efficiency} \label{eff}

Charged pions are identified by separating them from other charged hadrons (primarily charged kaons), and charged leptons (electrons and muons), as described in Sec.~II.A.  The efficiency for the $\pi/K$ separation criteria is studied using samples of data and MC, while the $\pi/e$ and $\pi/\mu$ separation efficiencies are directly determined from the data.

\textbf{Efficiency of $\bm{\pi/K}$ Separation}: \label{kaon}
The efficiency of the $\pi/K$ separation criteria for one track is determined as the fraction for the track to pass the kaon rejection criteria, while the other track is already selected as a pion or a muon. 
The average kaon rejection efficiency correction for one track of the two data sets is $(99.8\pm0.5)\%$.
Again, since there are two tracks in the event selections
we apply $C_{K}=[0.998(5)]^2=0.996(7)$ to the $\pi^+\pi^-$ cross section in this analysis. 

\begin{figure*}[!tb]
\begin{center}
\includegraphics[width=2.1in]{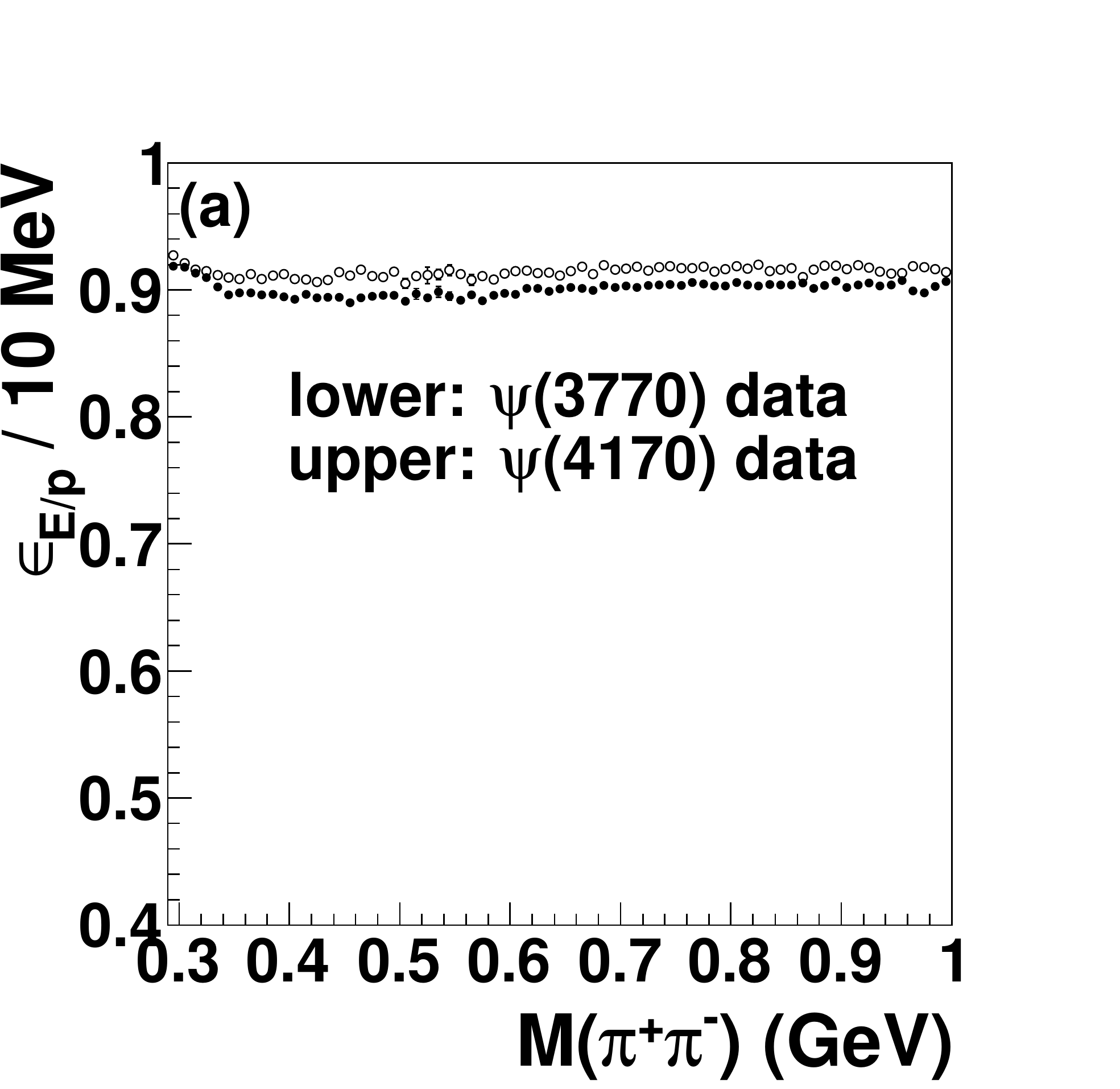}
\includegraphics[width=2.1in]{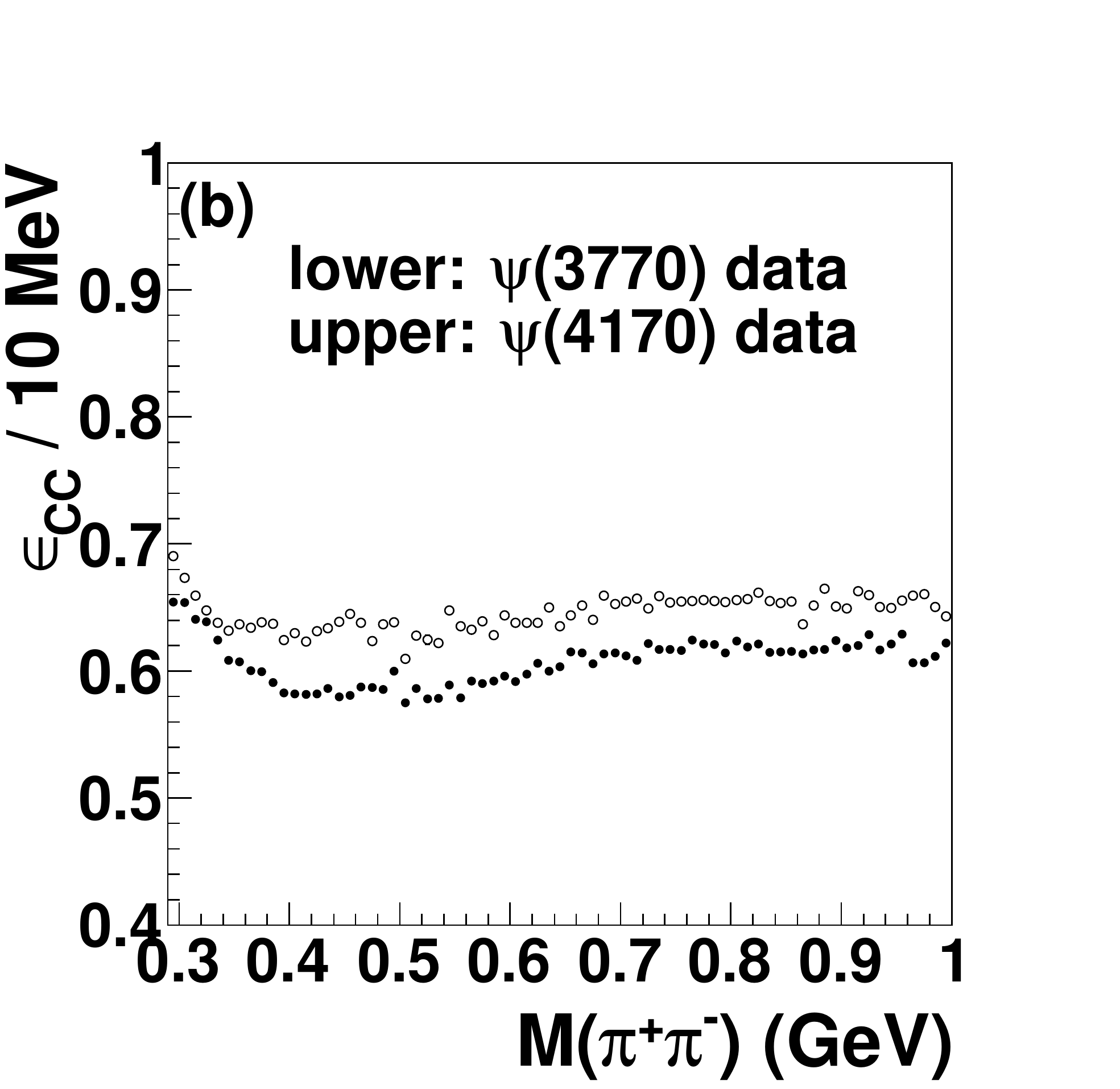}
\includegraphics[width=2.1in]{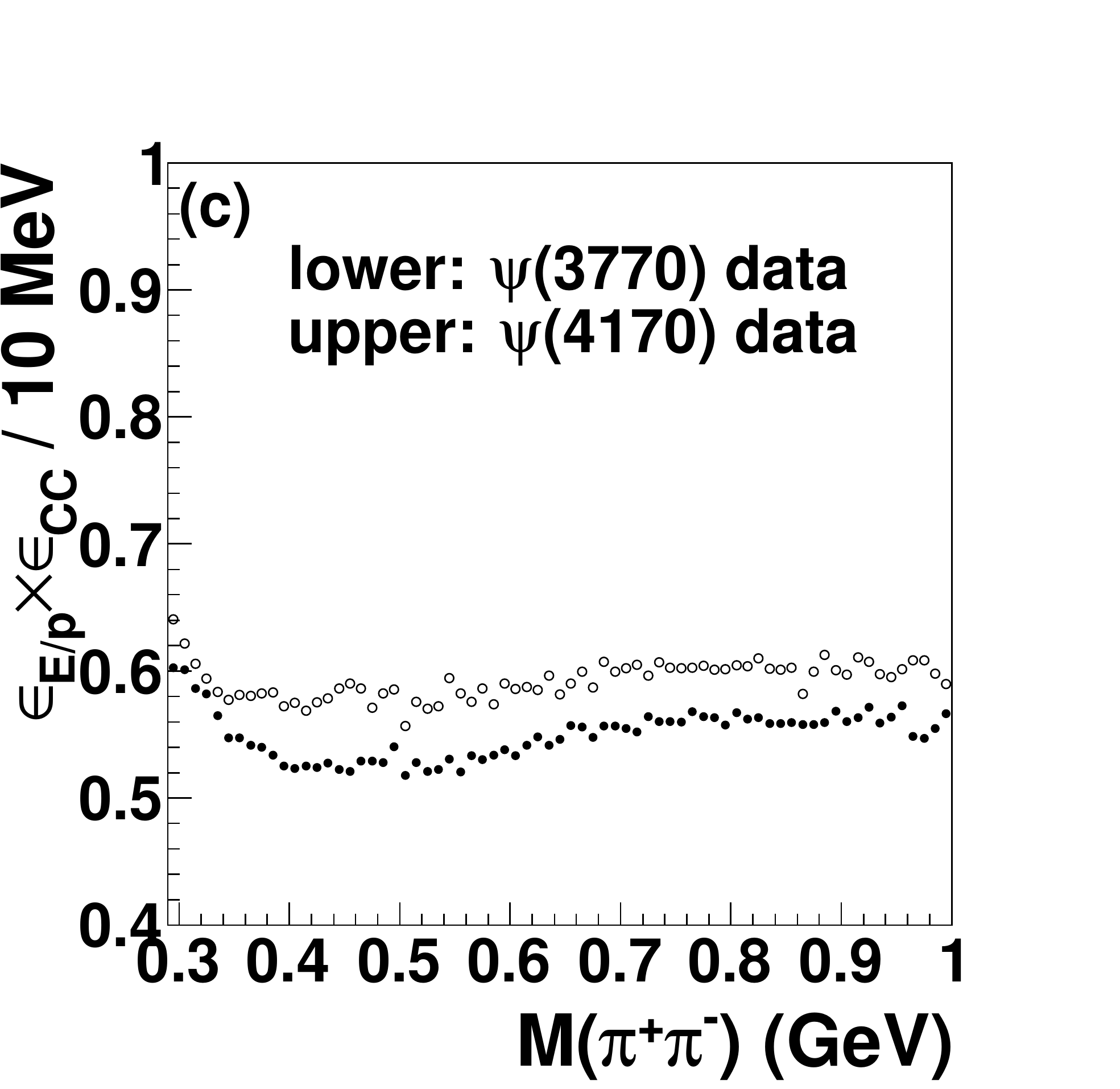}
\end{center}
\caption{(a) Efficiency of $\pi/e$ separation as a function of $M(\pi^+\pi^-)$. (b) Efficiency of $\pi/\mu$ separation as a function of $M(\pi^+\pi^-)$. (c)  The combined efficiency of $\pi/e$ and $\pi/\mu$ separations as a function of $M(\pi^+\pi^-)$.}
\label{fig:eff_cc_m}
\end{figure*}

\textbf{ Efficiency of $\bm{\pi/e}$ separation}:
As described in Sec.~II.A, we separate pions from electrons using the ratio $E_\mathrm{CC}/p$.  Since the nuclear interaction of pions in a calorimeter, and therefore the energy that they deposit, is difficult to accurately simulate, we determine $\epsilon_{\text{E}/p}$, the efficiency of $\pi/e$ separation using $\sim48~\mathrm{pb}^{-1}$ of data taken at the $\psi(2S)$ resonance. 
We first reject all but electron and pion tracks as follows: 
\begin{itemize}
\item events must have at least three tracks,
\item kaon-like or proton-like tracks are rejected using $dE/dx$ and $LL^{\mathrm{RICH}}$ as described in Sec.~\ref{pid},
\item tracks with hits in the muon chambers are rejected,
\item muons from the high-rate decay $\psi(2S)\to\pi^+\pi^-J/\psi$, $J/\psi\to\mu^+\mu^-$ are rejected as events with 3 or 4 total tracks and 2 tracks with a recoil mass within 5~MeV of $M(J/\psi)$.
\end{itemize}
The remaining tracks are mostly pions and electrons with electrons having $E_{\mathrm{CC}}/p>0.8$. To determine the efficiency of pions which survive the $E_{\mathrm{CC}}/p$ cut we fit the $E_{\mathrm{CC}}/p$ distributions separately for positive and negative tracks  in each 50~MeV momentum bin of our $\psi(2S)$ sample to separate the pion tail from the electron peak near  $E_{\mathrm{CC}}/p\sim 1$.
For all of our events, $\epsilon_{\text{E}/p}=\epsilon_{\text{E}/p}(\pi^+)\epsilon_{\text{E}/p}(\pi^-)$ is averaged in each bin of $M_{\pi\pi}$, as shown in Fig.~\ref{fig:eff_cc_m}(a).

\textbf{Efficiency of $\bm{\pi/\mu}$ separation}:
For $\pi/\mu$ separation we use the same $\psi(2S)$ data and criteria as described above for $\pi/e$ separation to determine the efficiency of the $E_{\mathrm{CC}}$-based muon rejection. The electrons are rejected using the criteria of $E_{\mathrm{CC}}/p > 0.8$.  The remaining tracks are mostly pions.

\begin{figure}[!tb]
\begin{center}
\includegraphics[width=3.2in]{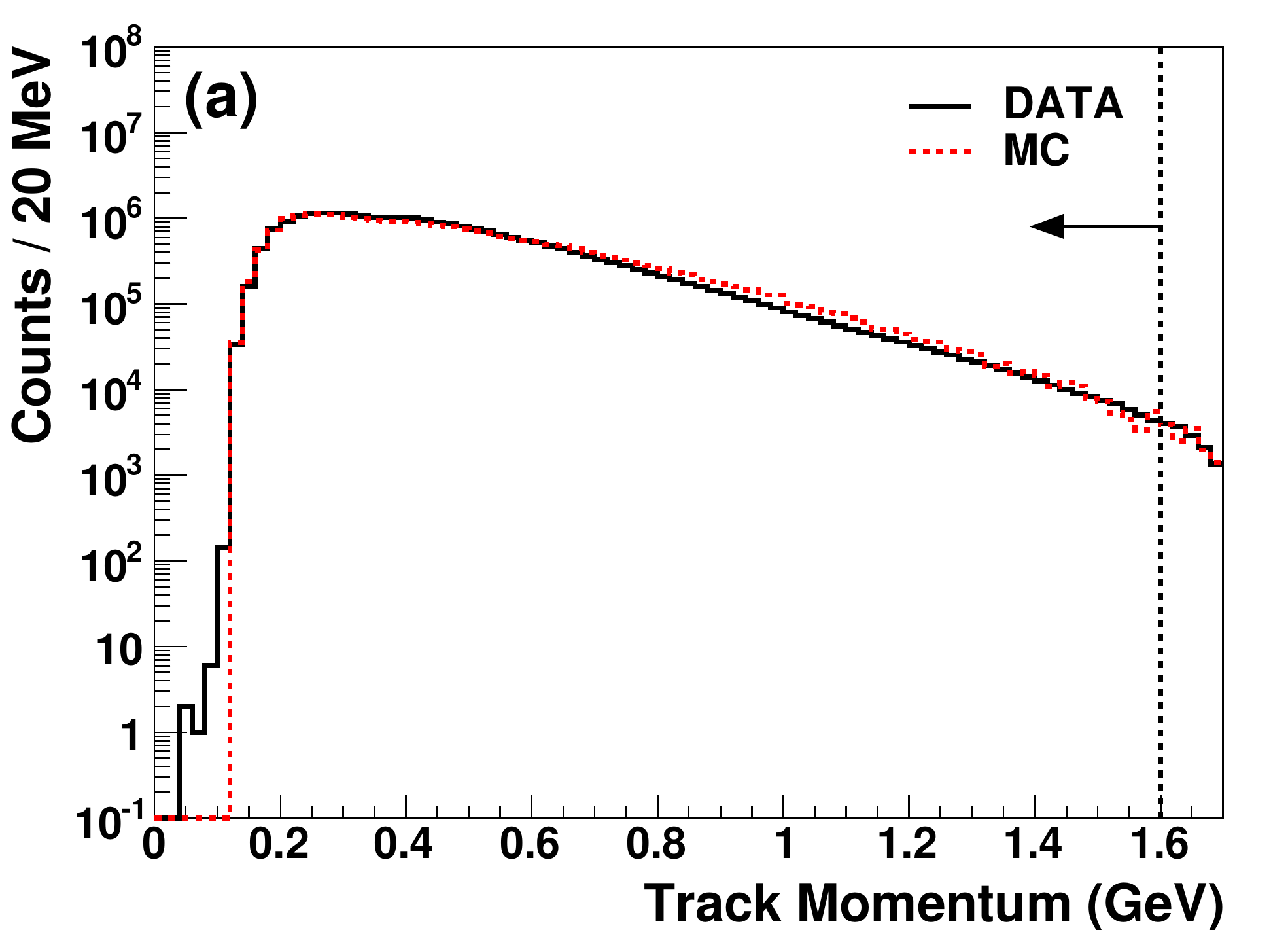}
\includegraphics[width=3.2in]{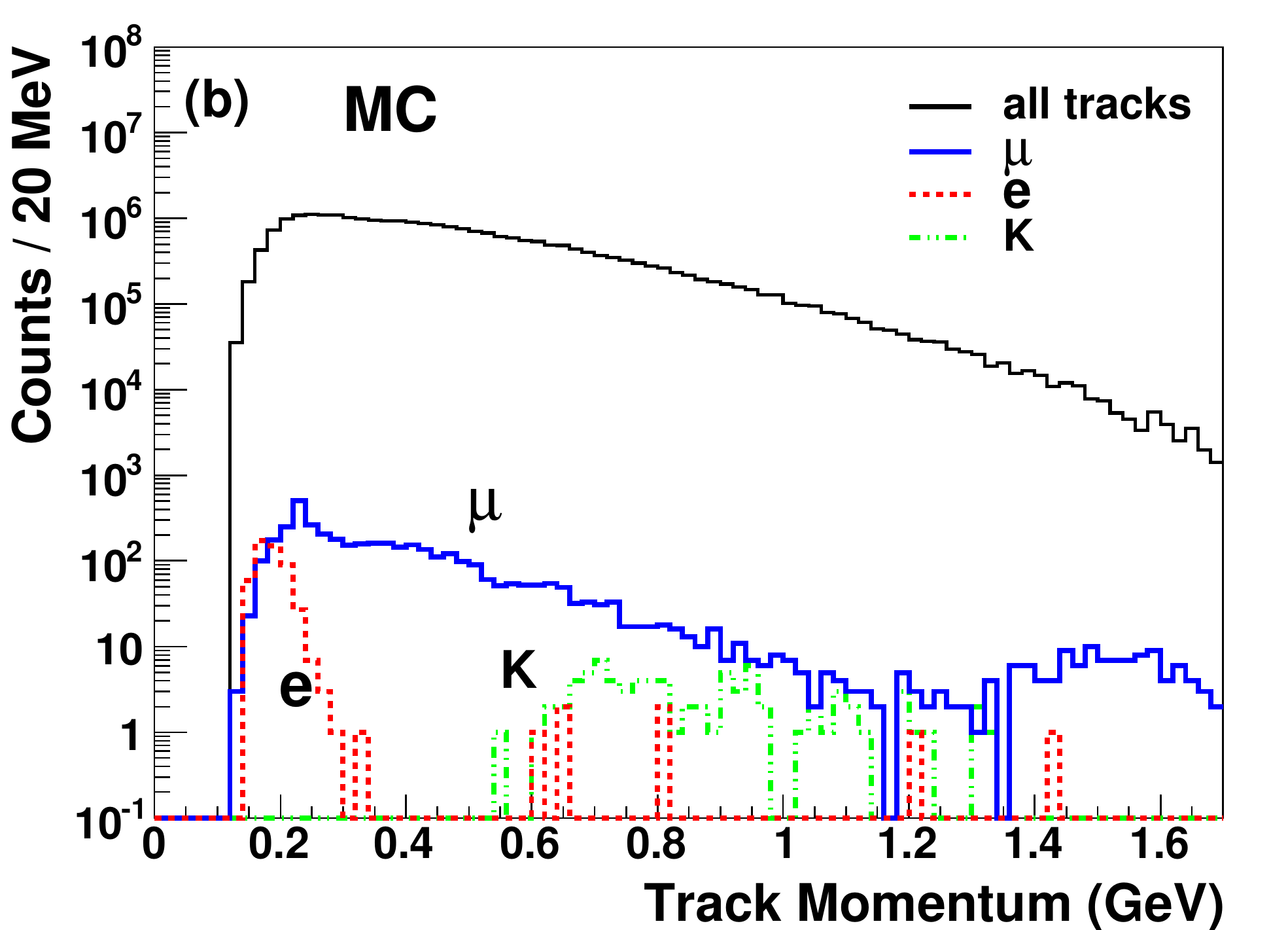}

\end{center}
\caption
{(a) Momentum distributions of all charged tracks from $\psi(2S)$ data and generic MC. The vertical dotted lines define the good statistics region in which data are accepted.
(b) Momentum distributions of MC-estimated individual contributions from muons, electrons, and kaons. The contribution from protons is too small to show in the plot.}
\label{fig:tr_mom_comp}
\end{figure}

For illustration, the momentum distributions of pions from $\psi(2S)$ data and generic MC are shown in Fig.~\ref{fig:tr_mom_comp}(a), together with the MC-estimated residual contributions from electrons, muons, kaons, and protons shown in Fig.~\ref{fig:tr_mom_comp}(b).  These particles are mostly pions. The estimated residual contributions from electrons, kaons and protons are negligible. The residual contribution from muons is  $\sim 0.02\%$.  We have compared the results with and without subtracting the residual contribution from muons using an estimate from simulated $\psi(2S)$ decays.  
The difference of $a_\mu(\pi^+\pi^-)$ is small $(\sim 0.1\%)$.

For each 50 MeV momentum bin in these data, we form $E_\text{CC}$ distributions for both positive and negative pion candidates. From these distributions we calculate the probability, $\epsilon_\text{CC} (\pi^\pm)$, for a single pion candidate to meet the muon-like criteria, $E_{\text{CC}}<0.3$~GeV. Thus for each event the efficiency of $\pi/\mu$ separation of pions from muons is:
\begin{equation}
\epsilon_\text{CC}  = 1 - \epsilon_\text{CC}(\pi^+)\epsilon_\text{CC} (\pi^-).
\end{equation}
These finial $\epsilon_\text{CC} $ efficiencies are averaged in each bin of $M(\pi\pi)$, as shown in Fig.~\ref{fig:eff_cc_m}(b).

\textbf{Analysis of $\bm{\psi(2S)\to\pi^+\pi^-J/\psi}$, $\bm{J/\psi\to\pi^+\pi^-\pi^0}$}:
Our determination of the $\pi/\mu$ separation efficiency $\epsilon_\text{CC}$  from $\psi(2S)$ inclusive decays can be confirmed by analysis of the exclusive decay
$\psi(2S)\to\pi^+\pi^-J/\psi$, $J/\psi\to\pi^+\pi^-\pi^0$. We reconstruct this decay by
selecting events with only four tracks and at least two showers that pass standard CLEO quality criteria for photons. The net charge of the tracks is required to be zero,
and at least one pair of photon candidates should form a $\pi^0$ candidate with mass within 15~MeV of the nominal $\pi^0$ mass of 135~MeV.
We obtain our $J/\psi$ sample from $\psi(2S)$ by tagging events with two oppositely charged tracks, which are assumed to be pions, and have a recoil mass in the range $M(J/\psi)\pm10$~MeV.
Events are rejected if any of the other two tracks from the $J/\psi$ decay is identified as a kaon or a proton in the same way described in Sec.~\ref{pid}.
These two tracks should also meet the requirements of $|\cos\theta|<0.75$ and $p_T>0.2$~GeV in the $M_{\pi\pi}$ region below 0.5~GeV.
We perform a 4C kinematic fit constraining the $2(\pi^+\pi^-)\pi^0$ to a common vertex with $\chi^2_{\mathrm{vertex}}<50$, and $e^+e^-$ collision energy and momentum with $\chi^2_{\mathrm{4C fit}}<50$.
We select pions from the decay of $J/\psi$ and compare the $E_\mathrm{CC}$ distributions with those determined from the inclusive decay of $\psi(2S)$ as described above.  The results are consistent within statistical uncertainty.

\section{Background Subtraction} \label{bkg}

After the cuts shown in the 2D plots of Fig.~\ref{fig:2Dfit_4170}, small amounts of backgrounds remain, particularly near the boundaries of the cuts.
To estimate these residual backgrounds, and to remove them from the $\pi^+\pi^-\gamma_{\mathrm{ISR}}$ sample, we estimate the contributions of individual background sources separately using data and MC samples in three different regions of $M_{\pi\pi}$: the low $M_{\pi\pi}$ range below $M(\rho)$ ($0.30-0.65$~GeV), the $\rho$ peak region ($0.65-0.85$~GeV), and the high $M_{\pi\pi}$ range above $M(\rho)$ ($0.85-1$~GeV). 
The results are presented in Table~\ref{tbl:background}. 

\begin{table*}[!tb]
\caption[Estimated background fractions (in $\%$) in the $\pi^+\pi^-\gamma_{\mathrm{ISR}}$ sample for the low $M_{\pi\pi}$ range ($0.30-0.65$~GeV), the $\rho$ peak ($0.65-0.85$~GeV), and the high $M_{\pi\pi}$ range ($0.85-1$~GeV).]
{Estimated background fractions (in $\%$) in the $\pi^+\pi^-\gamma_{\mathrm{ISR}}$ sample for the low $M_{\pi\pi}$ range ($0.30-0.65$~GeV), the $\rho$ peak ($0.65-0.85$~GeV), and the high $M_{\pi\pi}$ range ($0.85-1$~GeV). The first errors are statistical and the second errors are systematic.}
\begin{center}
\begin{tabular}{c|l|c|c|c}
\hline
 $~~\psi(3770)~~$   & ~~Bkg Source & Low $M_{\pi\pi}$ Range & $\rho$ Peak & High $M_{\pi\pi}$ Range \\ 
\hline
             & $~~q\bar{q}$ & $~~11.98\pm0.12\pm0.11~~$ & $~~3.21\pm0.04\pm0.03~~$ & $~~11.83\pm0.16\pm0.11~~$ \\
             & $~~\mu^+\mu^-\gamma_{\mathrm{ISR}}$ & $4.17\pm0.04\pm0.05$ & $0.37\pm0.01\pm0.01$ & $1.16\pm0.02\pm0.04$ \\                        
             & $~~\pi^+\pi^-\pi^0\gamma_{\mathrm{ISR}}$ & $2.87\pm0.03\pm0.09$ & $0.32\pm0.01\pm0.01$ & $0.50\pm0.01\pm0.02$ \\      
             &                                                                                          &                                              &                                             &                                              \\
             & $~~\pi^+\pi^-2\pi^0\gamma_{\mathrm{ISR}}~~$ & $0.06\pm0.01\pm0.01$ & $0.01\pm0.01\pm0.01$ & $0.04\pm0.01\pm0.01$ \\       
             & $~~K^+K^-\gamma_{\mathrm{ISR}}$ & $0.02\pm0.01\pm0.01$ & $<0.01$ & $<0.01$ \\ 
             & $~~e^+e^-\gamma$ & $<0.01$ & $<0.01$ & $<0.01$ \\
             & $~~J/\psi\gamma_{\mathrm{ISR}}$ & $0.10\pm0.02\pm0.01$ & $0.15\pm0.01\pm0.01$ & $0.30\pm0.04\pm0.01$ \\
             & $~~\psi(2S)\gamma_{\mathrm{ISR}}$ & $0.05\pm0.01\pm0.01$ & $0.03\pm0.01\pm0.01$ & $0.07\pm0.02\pm0.01$ \\   
\hline
             & ~~total &        $19.25\pm0.13\pm0.15$    & $4.09\pm0.05\pm0.04$ & $13.90\pm0.17\pm0.12$ \\  
\hline\hline

$~~\psi(4170)~~$ & $~~q\bar{q}$ & $5.75\pm0.10\pm0.07$ & $1.92\pm0.03\pm0.02$ & $7.12\pm0.15\pm0.09$ \\
             & $~~\mu^+\mu^-\gamma_{\mathrm{ISR}}$ & $4.21\pm0.05\pm0.06$ & $0.28\pm0.01\pm0.01$ & $1.01\pm0.03\pm0.04$ \\                        
             & $~~\pi^+\pi^-\pi^0\gamma_{\mathrm{ISR}}$ & $3.23\pm0.04\pm0.11$ & $0.34\pm0.01\pm0.01$ & $0.56\pm0.02\pm0.02$ \\    
             &                                                                                          &                                              &                                             &                                              \\
             & $~~\pi^+\pi^-2\pi^0\gamma_{\mathrm{ISR}}~~$ & $0.12\pm0.02\pm0.01$ & $0.02\pm0.01\pm0.01$ & $0.04\pm0.02\pm0.01$ \\    
             & $~~K^+K^-\gamma_{\mathrm{ISR}}$ & $0.02\pm0.01\pm0.01$ & $<0.01$ & $<0.01$ \\ 
             & $~~e^+e^-\gamma$ & $<0.01$ & $<0.01$ & $<0.01$ \\
             & $~~J/\psi\gamma_{\mathrm{ISR}}$ & $0.06\pm0.01\pm0.01$ & $0.12\pm0.01\pm0.01$ & $0.26\pm0.04\pm0.01$ \\
             & $~~\psi(2S)\gamma_{\mathrm{ISR}}$ & $<0.01$ & $<0.01$ & $<0.01$ \\      
\hline
             & ~~total &        $13.39\pm0.12\pm0.14$    & $2.68\pm0.04\pm0.03$ & $8.99\pm0.16\pm0.10$ \\  
\hline
\end{tabular}
\end{center}
\label{tbl:background}
\end{table*}

\begin{enumerate}
\item Background from $e^+e^-\to q\bar{q}\to\text{light~hadrons}$ arises from those decays in which $\pi^0$'s are produced, and photons from $\pi^0\to\gamma\gamma$ decays are mistaken as ISR photon candidates. 
This background is considerably reduced by the kinematic fit $\chi^2$ selections and the $\pi^0$ veto cut.
The contribution of this background in our analysis is estimated using simulated samples of the $e^+e^-\to q\bar{q}\to\text{light~hadrons}$ process.
To avoid additional uncertainties in the cross section for $e^+e^-\to q\bar{q}\to\text{light~hadrons}$, we normalize the simulated $q\bar{q}$ sample to our data in the following way. 
In the background region of the 2D $\chi^2$ plot, i.e. for events in the large box, we pair the primary ISR photon candidate with all detected additional photons and keep the pair with $\gamma\gamma$ mass closest to the nominal $\pi^0$ mass of 135~MeV. 
The normalization factor $f=N(\text{data})/N(\text{MC})$ is then obtained by fitting the $\pi^0\to\gamma\gamma$ yield in data and in the simulation, as shown in Table~\ref{tbl:scaling}.

\item We find that three processes of the type $e^+e^-\to \gamma_\text{ISR}+\text{light~hadrons}$ contribute non-negligible amounts to the background.  Their contributions are estimated as follows using MC-simulated events in a similar way to the $q\bar{q}$ process. 
\begin{itemize}
\item For the $K^+K^-$ ISR process, we normalize the MC yield to data by fitting the $\phi$ peak.
\item For the $\pi^+\pi^-\pi^0$ ISR process, we normalize the MC yield to data using the yields of $\omega$ and $\phi$ resonances. 
\item For the $\pi^+\pi^-2\pi^0$ ISR process, we estimate the normalization factor by comparing the total yields at $1<M(\pi^+\pi^-2\pi^0)<2$~GeV in data and MC.
\end{itemize}
And the resulting scale factors are listed in Table~\ref{tbl:scaling}.

\item Background from $e^+e^-\to\psi(1S,2S)\gamma_{\mathrm{ISR}}$ is estimated from MC.
The scale factors are estimated as the ratio of the number of the expected events and of the generated events. They are also listed in Table~\ref{tbl:scaling}.

\begin{table}[!tb]
\caption
{Scale factors normalized from data for background processes.}
\begin{center}
\begin{tabular}{clc}
\hline
    & Bkg Source  & Scale Factor \\ 
\hline
$\psi(3770)$ & $q\bar{q}$ & $0.0434\pm0.0004$  \\   
             & $K^+K^-\gamma_{\mathrm{ISR}}$ & $0.0128\pm0.0003$  \\              
             & $\pi^+\pi^-\pi^0\gamma_{\mathrm{ISR}}$  & $0.0092\pm0.0003$ \\      
             & $\pi^+\pi^-2\pi^0\gamma_{\mathrm{ISR}}$ & $0.0697\pm0.0020$  \\    
             & $J/\psi\gamma_{\mathrm{ISR}}$ & $0.1305\pm0.0002$  \\
             & $\psi(2S)\gamma_{\mathrm{ISR}}$ & $0.1373\pm0.0001$  \\      

\hline
$\psi(4170)$ & $q\bar{q}$ & $0.0329\pm0.0004$  \\   
             & $K^+K^-\gamma_{\mathrm{ISR}}$ & $0.0098\pm0.0003$  \\              
             & $\pi^+\pi^-\pi^0\gamma_{\mathrm{ISR}}$  & $0.0089\pm0.0003$ \\      
             & $\pi^+\pi^-2\pi^0\gamma_{\mathrm{ISR}}$ & $0.0871\pm0.0019$  \\    
             & $J/\psi\gamma_{\mathrm{ISR}}$ &  $0.0505\pm0.0001$ \\
             & $\psi(2S)\gamma_{\mathrm{ISR}}$ & $0.0378\pm0.0001$  \\  

\hline
\end{tabular}
\end{center}
\label{tbl:scaling}
\end{table}

\item Radiative Bhabha events ($e^+e^-\to e^+e^-\gamma$) are very strongly suppressed by the requirement $E_\mathrm{CC}/p>0.8$. The number of remaining events due to this process is estimated as the production of the expected number of radiative Bhabha events and the efficiency for these events to pass our event selections.
The efficiency for radiative Bhabha events to pass the electron rejection cut is estimated using data.
We reconstruct events with the ISR photon candidate and two oppositely-charged tracks, requiring that one track is an electron ($0.9<E_\mathrm{CC}/p<1.1$), and determine the probability that the other track would pass our electron rejection cut. This gives the probability for a single electron. Assuming the two electrons are uncorrelated, we get the total efficiency by squaring the one-electron probability. 
The efficiency for radiative Bhabha events to pass our event selection criteria except the electron rejection cut is estimated using simulated radiative Bhabha events.
The estimated number of radiative Bhabha events is below the $10^{-4}$ level of the total background.

\item Background processes $p\bar{p}\gamma_{\mathrm{ISR}}$ and $\tau^+\tau^-$ contribute significantly only at $\pi^+\pi^-$ masses much higher than the range of interest for the present analysis. We estimate the contributions of these channels using simulated events generated by EvtGen. Neither exceeds the $10^{-4}$ level in the fraction of the total background.
\end{enumerate}

\begin{figure}[!tb]
\begin{center}

\includegraphics[width=3.2in]{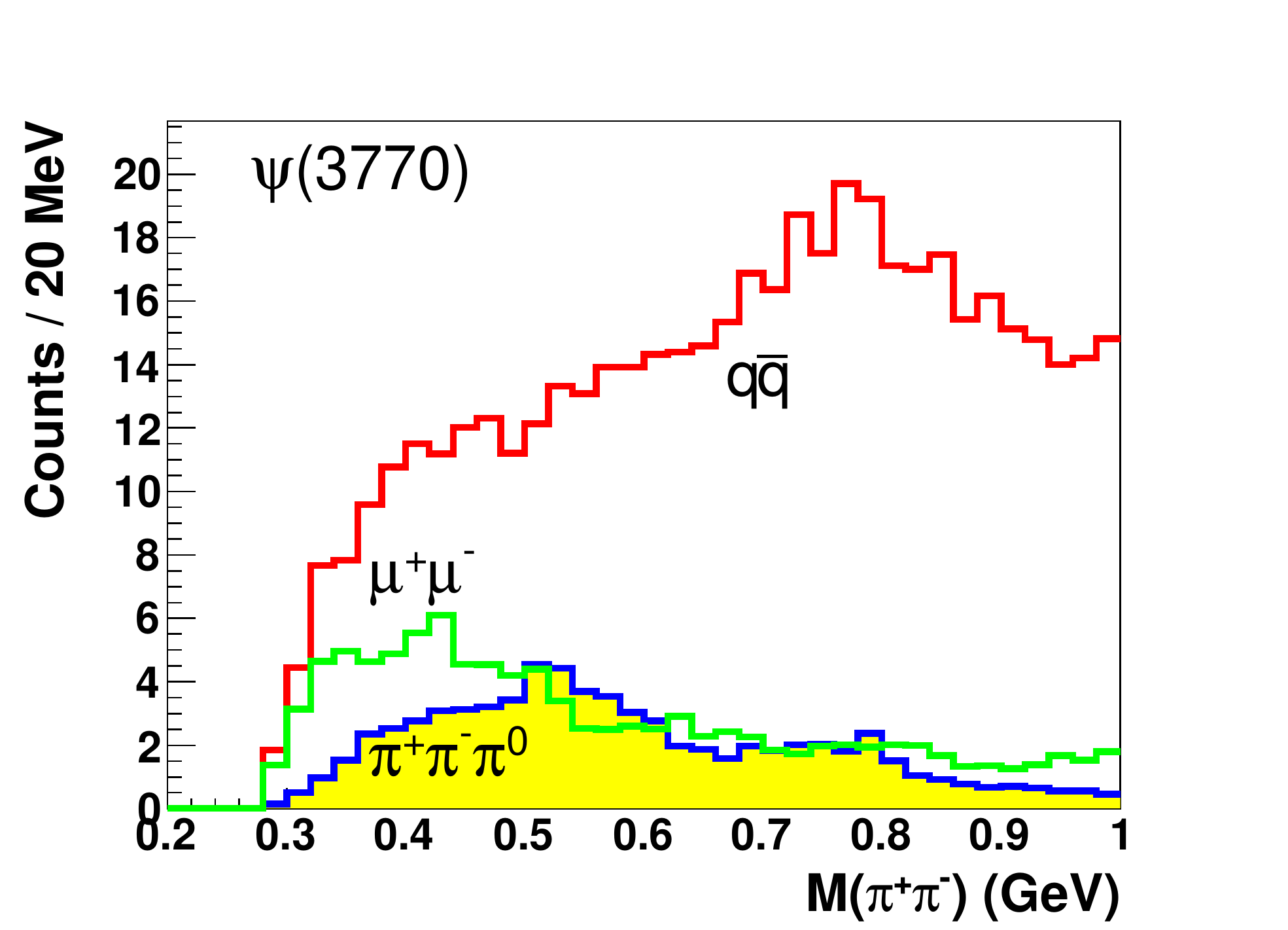}
\includegraphics[width=3.2in]{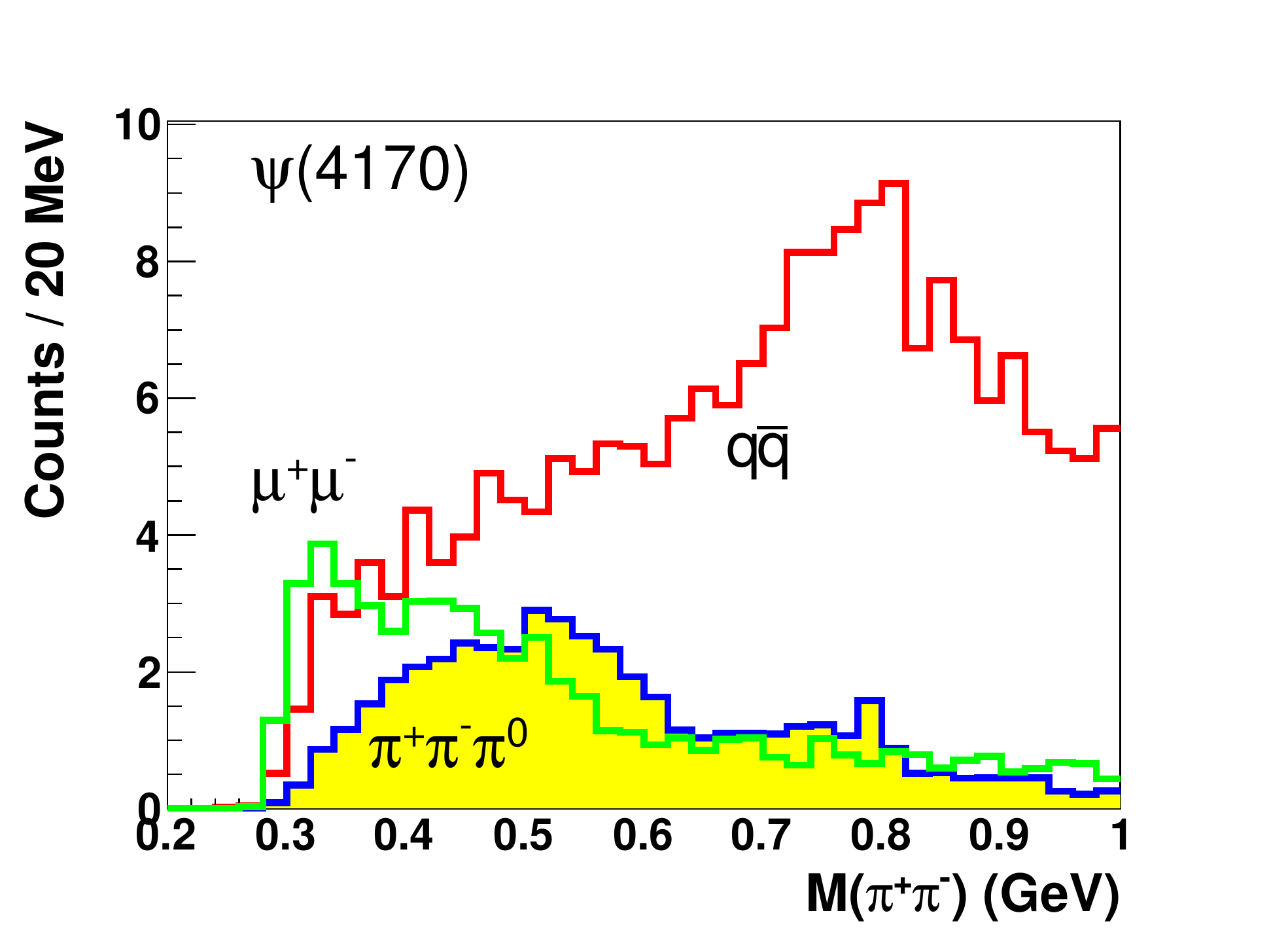}

\end{center}
\caption
{Contributions of the background from $e^+e^-\to q\bar{q}$, $e^+e^-\to\gamma_{\mathrm{ISR}}\pi^+\pi^-\pi^0$, and $e^+e^-\to\mu^+\mu^-\gamma_{\mathrm{ISR}}$ in the $M_{\pi\pi}$ distribution. Contributions of other backgrounds are too small to show in the plots. }
\label{fig:mmpi_bkg}
\end{figure}

As we note in Table~\ref{tbl:background}, the main background processes are $q\bar{q}$, $\mu^+\mu^-\gamma_{\mathrm{ISR}}$ and $\pi^+\pi^-\pi^0\gamma_{\mathrm{ISR}}$.
Fig.~\ref{fig:mmpi_bkg} shows contributions from these main background processes.

\begin{figure}[!tb]
\begin{center}
\includegraphics[width=3.3in]{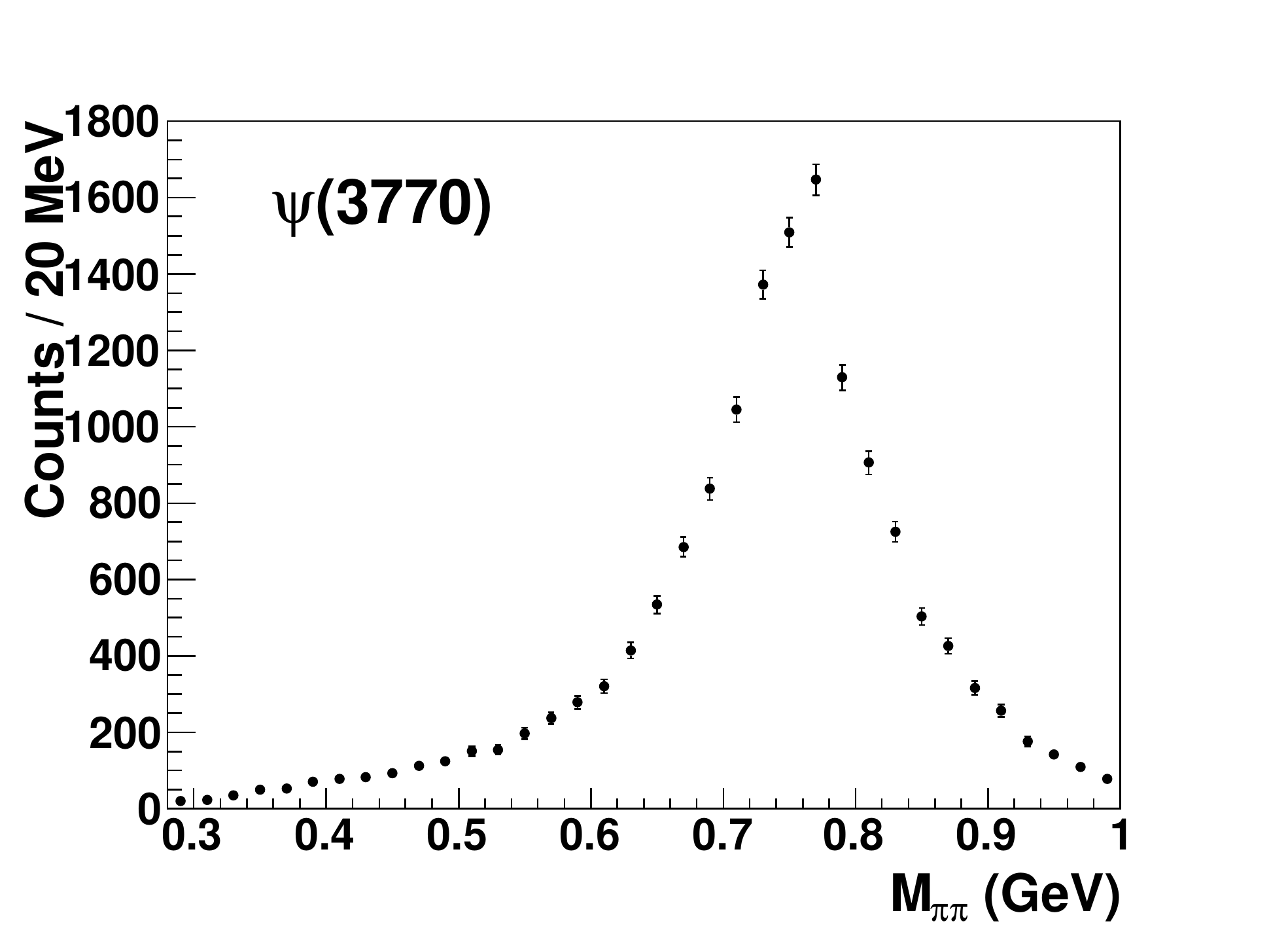}
\includegraphics[width=3.3in]{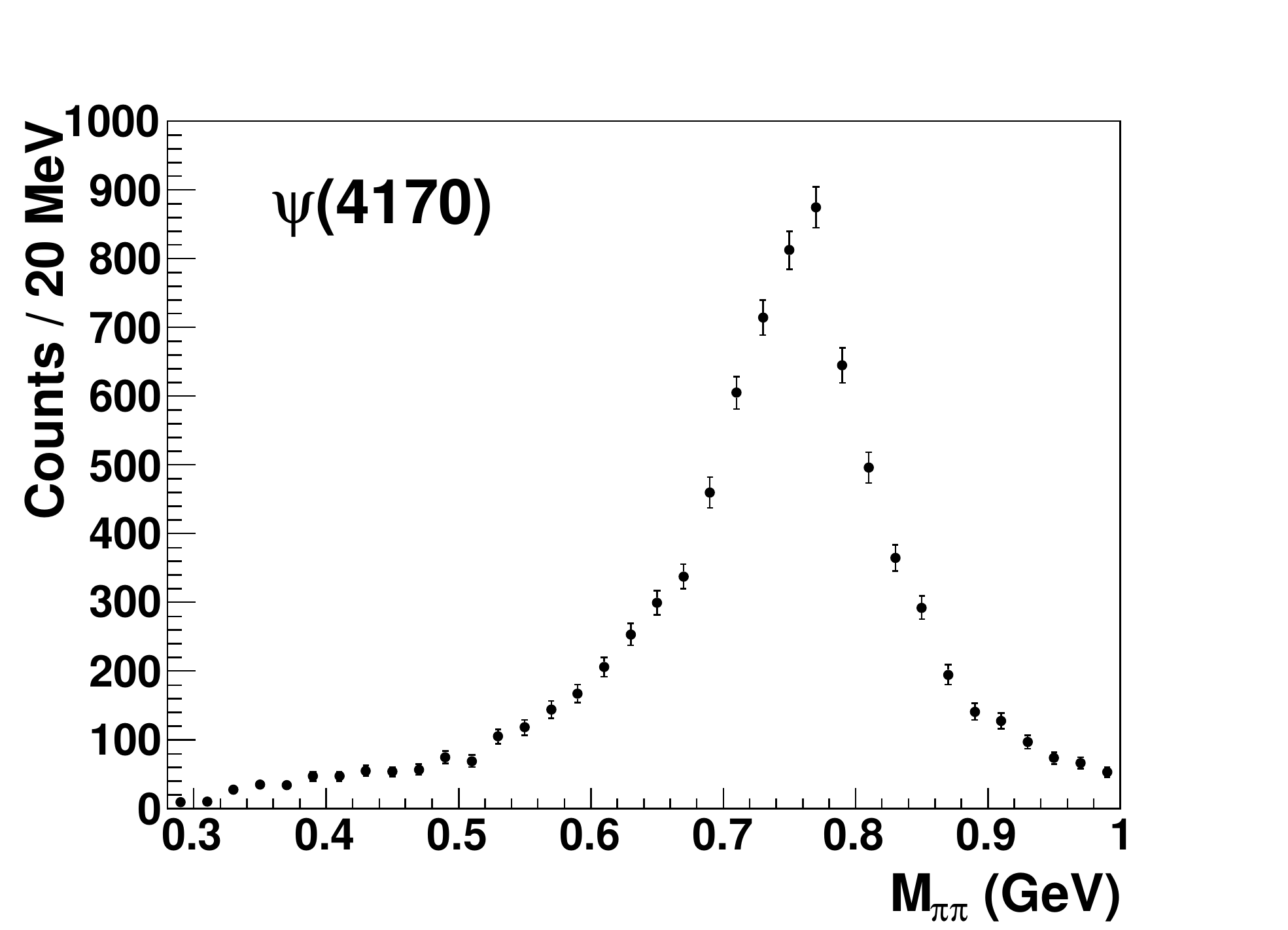}
\end{center}
\caption
{The invariant mass distribution $M_{\pi\pi}$ for $\psi(3770)$ and $\psi(4170)$ data sets.}
\label{fig:dist}
\end{figure}

After applying the event selections described in Sec.~II and rejecting all other backgrounds, we determine that the event sample only consists of $\pi^+\pi^-\gamma_{\mathrm{ISR}}$ signal events and $\sim1.5\%$ residual $\mu^+\mu^-\gamma_{\mathrm{ISR}}$ background events.
We separate the $\mu^+\mu^-\gamma_{\mathrm{ISR}}$ events from the final $\pi^+\pi^-\gamma_{\mathrm{ISR}}$ sample in each bin by the following procedure:

We denote the numbers of $\pi^+\pi^-$ and $\mu^+\mu^-$ produced as $N^{(0)}_{\pi\pi}$ and $N^{(0)}_{\mu\mu}$ respectively, and as $N_{\pi\pi}$ and $N_{\mu\mu}$ as the measured numbers after the $\pi/\mu$ separation criteria are applied. 
We denote by $\epsilon_{\mathrm{CC}}^{\pi\pi}$ the efficiency for pions to have survived this separation procedure using the $\psi(2S)$ data as described in Sec.~\ref{eff}. 
We denote by $\epsilon_{\mathrm{CC}}^{\mu\mu}$ the efficiency for pions to have survived this procedure using $e^+e^-\to\mu^+\mu^-\gamma_{\mathrm{ISR}}$ events generated by Phokhara.
The numbers of pions and muons produced, $N^{(0)}_{\pi\pi}$ and $N^{(0)}_{\mu\mu}$ are then obtained by solving the equations:
\begin{equation}
 \begin{array}{ll}
 N_{\pi\pi}=N^{(0)}_{\pi\pi}\epsilon^{\pi\pi}_\mathrm{CC}+N^{(0)}_{\mu\mu}\epsilon^{\mu\mu}_\mathrm{CC}  \\
 N_{\mu\mu}=N^{(0)}_{\pi\pi}(1-\epsilon^{\pi\pi}_\mathrm{CC})+N^{(0)}_{\mu\mu}(1-\epsilon^{\mu\mu}_\mathrm{CC})
       \end{array} 
\end{equation} 
Thus, the number of these $\pi^+\pi^-\gamma_{\mathrm{ISR}}$ events which pass all our event selections is $N^{(0)}_{\pi\pi}\epsilon^{\pi\pi}_\mathrm{CC}$. The invariant mass distributions for $\pi^+\pi^-$ are shown in Fig.~\ref{fig:dist} for the $\psi(3770)$ and the $\psi(4170)$ data sets.
Because of the relatively large bin size our $M_{\pi\pi}$ spectra, unfolding procedures do not affect the final result of $a_\mu^{\pi\pi}$. Therefore, we do not apply an unfolding procedure to our $M_{\pi\pi}$ distributions.

\begin{figure}[!tb]
\begin{center}
\includegraphics[width=3.3in]{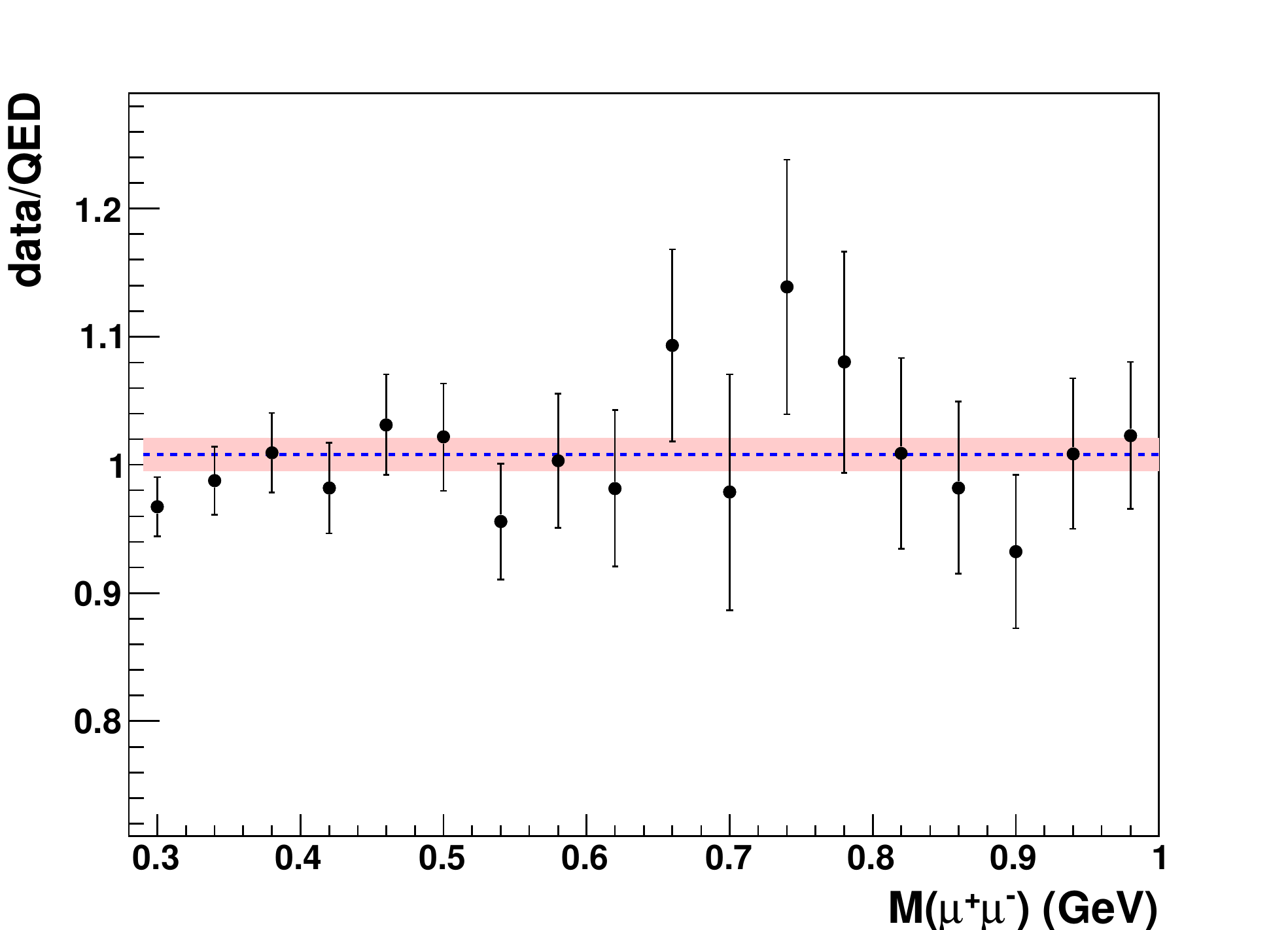}
\end{center}
\caption[]{The ratio of the invariant $\mu^+\mu^-$ mass spectrum in data over the NLO QED prediction from Phokhara using the luminosities of the $\psi(3770)$ and $\psi(4170)$ data sets.}
\label{fig:mumu}
\end{figure}

The validity of our $\pi/\mu$ separation procedure is confirmed by the general agreement of the observed $\mu^+\mu^-$ event distribution with the QED prediction for the same distribution.
The number of $\mu^+\mu^-\gamma_{\mathrm{ISR}}$ events which are rejected by the $\pi/\mu$ separation procedure is $N^{(0)}_{\mu\mu}(1-\epsilon^{\mu\mu}_\mathrm{CC})$.
The NLO QED prediction for the invariant $\mu^+\mu^-$ mass spectrum is obtained from the Phokhara generator, normalized to the data luminosity.
Fig.~\ref{fig:mumu} shows the ratio of the invariant $\mu^+\mu^-$ mass spectrum in data and the QED prediction. 
Good agreement is found  at the level of $(0.8\pm1.3)\%$.

\begin{figure}[!tb]
\begin{center}
\includegraphics[width=3.2in]{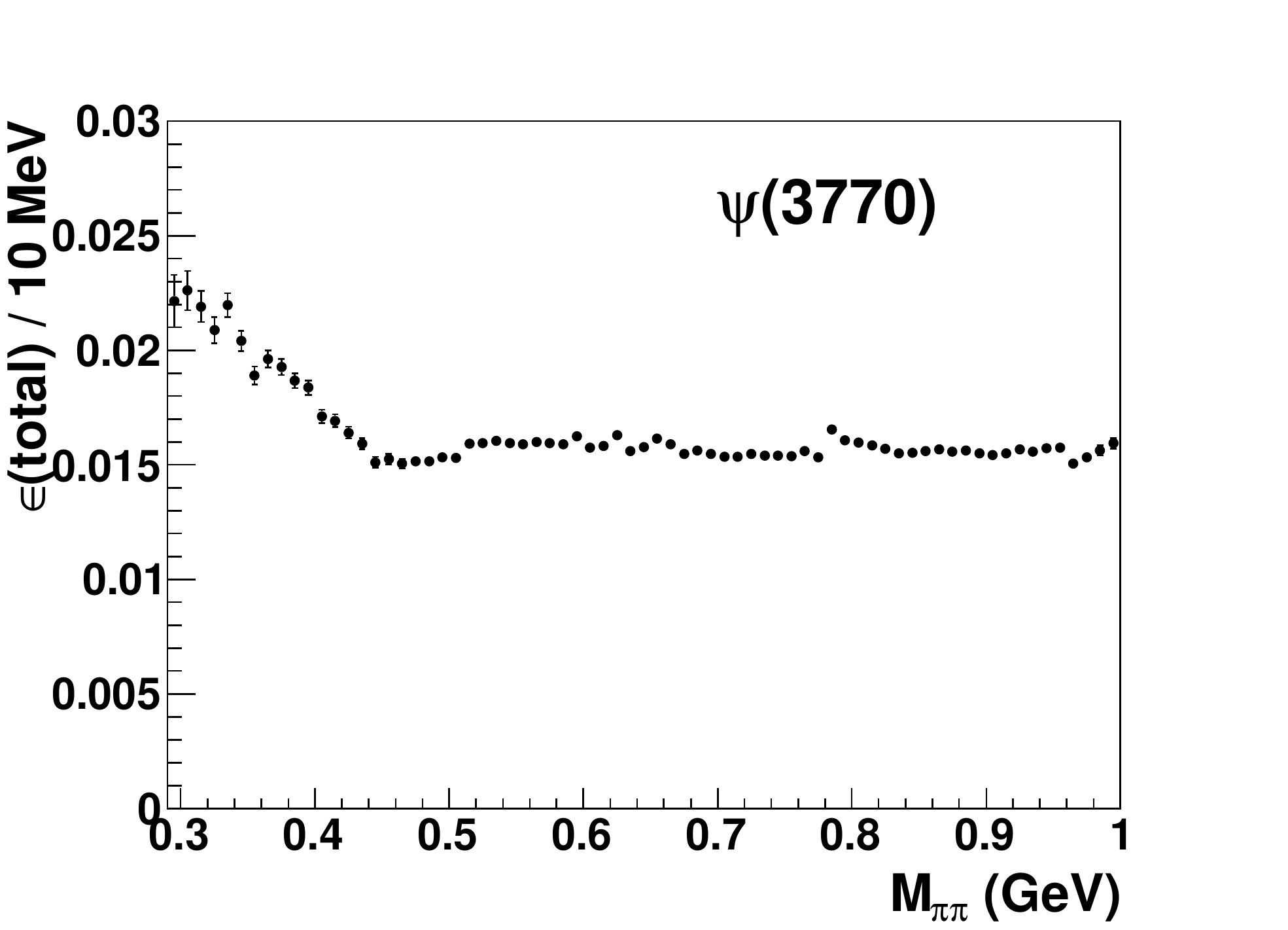}
\includegraphics[width=3.2in]{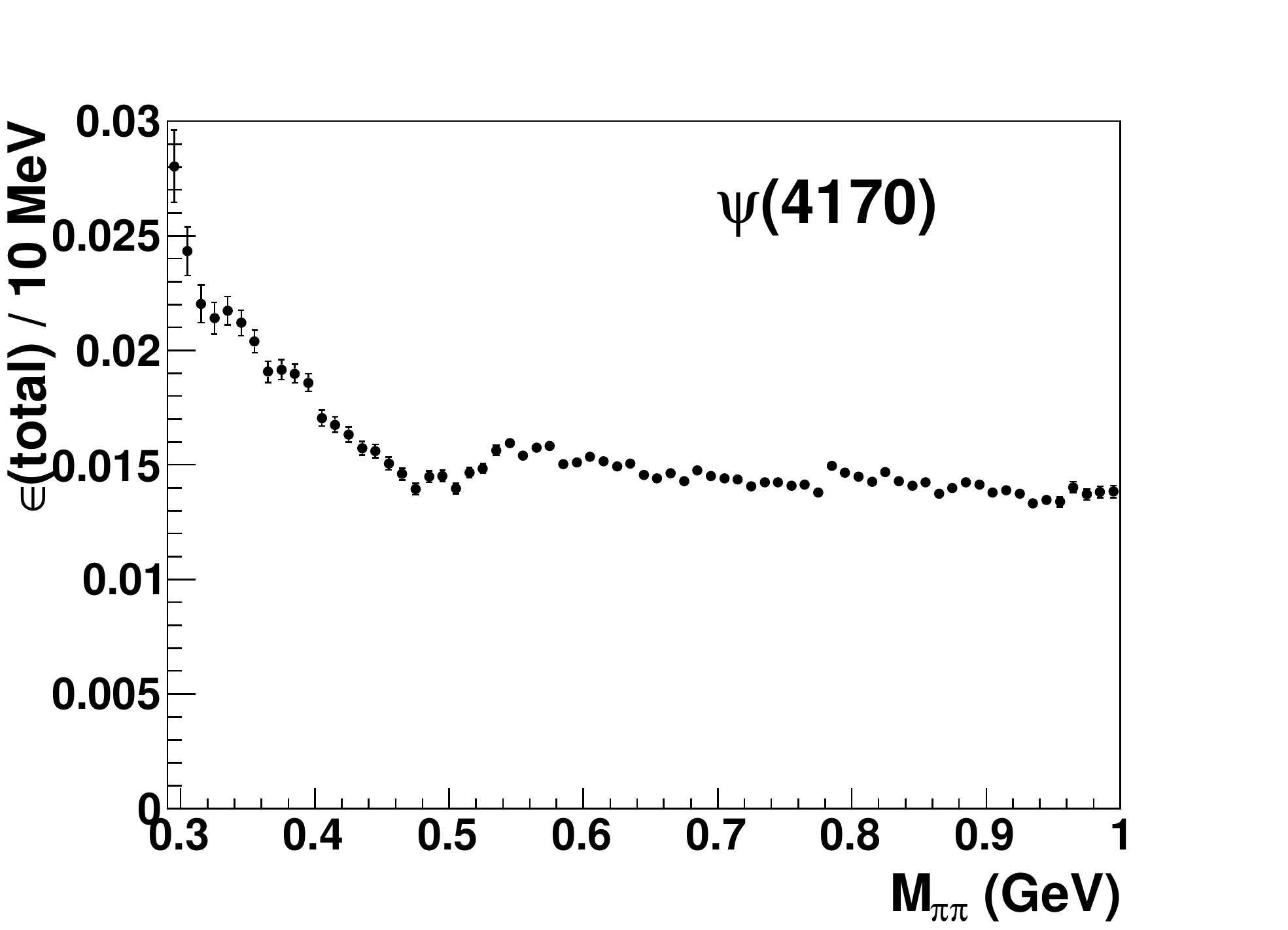}
\end{center}
\caption[The total efficiency to select $\pi^+\pi^-\gamma_{\mathrm{ISR}}$ events as a function of $M_{\pi\pi}$.]
{The total efficiency to select $\pi^+\pi^-\gamma_{\mathrm{ISR}}$ events as a function of $M_{\pi\pi}$. The small discontinuity near the $\rho-\omega$ interference region is caused by resolution effects.
The dip at $M_{\pi\pi}\approx0.5$~GeV region is due to the additional $p_T>0.2$~GeV requirement for tracks in the $M_{\pi\pi}$ region below 0.5~GeV,
and the rejection of the events in which two tracks merge in the CC and produce overlapping $E_\mathrm{CC}$.}
\label{fig:eff_total}
\end{figure}

Fig.~\ref{fig:eff_total} shows the total efficiency of our final event selection.
Contributions from individual sources are listed in Table~\ref{tbl:eff_list}.

\begin{table*}[!tb]
\caption{Efficiencies for selecting the $\pi^+\pi^-(\gamma)\gamma_{\mathrm{ISR}}$ events.}
\begin{center}
\begin{tabular}{|l|c|c|c|}
\hline
Cuts & Requirements & $\epsilon(\psi(3770)) (\%)$ & $\epsilon(\psi(4170)) (\%)$ \\
\hline
             & Standard CLEO quality criteria  &   &\\
             & $N_{\mathrm{trk}}=2$  &  & \\
Acceptance             & $\sum Q=0$                      & 10.93 & 10.53\\
             & $E_{\mathrm{ISR}}>50$~MeV   &   &  \\
             & E9/E25 cut on $\gamma_{\mathrm{ISR}}$  &  & \\
\hline
$\pi^0$ rejection & $|M_{\gamma\gamma}-M_{\pi^0}|>20$~MeV   &97.33  & 97.98 \\
\hline
$|\cos\theta|$ of tracks  & $|\cos\theta_{\mathrm{trk}}|<0.75$  & 50.59  & 51.41 \\
\hline
$p_T$ of tracks  &  $p_T>0.2$~GeV for low  $M_{\pi\pi}$ region & 99.87  & 99.92 \\
\hline
$|p|$ of tracks  &  $|p|<1.6$~GeV & 90.49  & 76.97 \\
\hline
1D $\chi^2$ & $\ln(\chi^2+1)<6$ &96.84 & 96.83 \\
\hline
2D $\chi^2$ & Signal region definition &70.18 & 71.02 \\
\hline
Overlapping tracks & Reject overlapping tracks in the CC &99.33 & 99.21 \\
\hline
$K$ rejection & $\Delta L_{K,\pi}>0$ & 97.49 & 97.09 \\
\hline
$e$ rejection & $E_{\mathrm{CC}}/p<0.8$ & 92.69 & 93.91 \\
\hline
$\mu$ rejection & $E_{\mathrm{CC}}>0.3$~GeV & 59.94  & 63.90 \\
\hline
bkg rejection  & Residual bkg rejection &91.69 & 94.45\\
\hline
Total  &  &1.63 & 1.53\\
\hline

\end{tabular}
\end{center}
\label{tbl:eff_list}
\end{table*}

\section{Systematic Uncertainties} \label{sys}

We estimate systematic uncertainties in $\sigma_{\pi\pi}$ and $a^{\pi\pi}_\mu$ for the combined results for the $\psi(3770)$, $\sqrt{s}=3770$~MeV, and $\psi(4170)$, $\sqrt{s}=4170$~MeV data sets.  As described in previous sections, we evaluate corrections to our efficiency determination using data where possible. As it is not possible to evaluate systematic uncertainties reliably in regions of small event statistics, we evaluate the systematic uncertainties due to each source for the full range of $M(\pi\pi)$ wherever possible.
These systematic uncertainties described in detail below and are summarized in Table~\ref{tbl:sys1}.

Efficiency corrections for triggering, tracks reconstruction, ISR photon reconstruction and $\pi/K$ separation have been described in Section~\ref{efficiency}. 
The finite statistical precision in these efficiency determinations is taken to be the corresponding systematic uncertainties due to these sources.

The efficiencies for $\pi/e$ and $\pi/\mu$ separations are determined using $\psi(2S)$ data, as described in Sections~\ref{eff}.  The systematic uncertainty in the determination of these efficiencies can arise from the following sources:
\begin{enumerate}
\item \textit{Bin size}: The efficiencies are determined in 50~MeV bins of momentum.  To estimate the effect of this choice of bin size, we reanalyze the data using 20~MeV bins as well.
\item \textit{Muon subtraction}: We have subtracted the contribution from muon tracks in these efficiency determinations as shown in Fig.~\ref{fig:tr_mom_comp}. Since this muon track background only contributes to $\sim 0.02\%$ of the total tracks, we determine this contribution by determining $a_\mu^{\pi\pi}$ without subtracting it. 
\item \textit{Electron subtraction}: The contribution from electron tracks is obtained by fitting the peak at $E_{\mathrm{CC}}/p\sim 1$ in the $E_{\mathrm{CC}}/p$ distribution of the tracks. To determine the systematic uncertainty in this yield determination, we vary the fit range and the order of the polynomial fit. The largest change in resulting $a_\mu^{\pi\pi}$ is taken as the systematic uncertainty due to this source. 
\item \textit{Finite statistics}: We include the statistical uncertainty in the determination of these efficiencies from $\psi(2S)$ data. 
\end{enumerate}
These individual contributions are all $\sim 0.1\%$ and their sum in quadrature is taken to be the systematic uncertainty due to the efficiency determination for $\pi/e$ and $\pi/\mu$ separation.

The systematic uncertainties due to the determination of the the charged particle momentum scale and uncertainties in the angular resolution are estimated by systematically varying these values by $\pm1\sigma$.  An estimation of the systematic error in the absolute energy scale, separately in the CC barrel and endcap, is taken into account to estimate the systematic uncertainty due to this source.

The systematic uncertainty due to the background rejection method is estimated by varying the normalization scale factor in each background channel by the of its uncertainty. The largest deviation is taken as the systematic uncertainty due to this source. 

The systematic uncertainty due to $\pi^0$ rejection is estimated by varying the $\pi^0$ rejection cut by 5~MeV.  The largest deviations are selected as the systematic uncertainty due to $\pi^0$ rejection.  Similarly, for kinematic fit we vary $\mathrm{ln}(\chi^2+1)$ around the nominal values by 2 for 1D distribution and by 0.5 for 2D distribution. The largest deviations are selected as the systematic uncertainties due to kinematic fit $\chi^2$ selections.

The systematic uncertainty due to the $e^+e^-$ luminosity is determined to be $0.6\%$ in Ref.~\cite{luminosity}.

\begin{table}[!tb]
\caption{Systematic uncertainties on $a_\mu(\pi^+\pi^-)$.}
\begin{center}
\begin{tabular}{lc}
\hline
Source  &  Systematic uncertainty (\%)  \\
\hline
$C_{\mathrm{trig}}$  &  0.5           \\
$C_{\mathrm{track}}$   &     0.6  \\
$C_{\gamma}$   &  0.6     \\
$C_{K}$       &      0.7 \\
$\epsilon_{\mathrm{CC}}$ and $\epsilon_{\mathrm{E/p}}$ determination  & 0.2\\
Tracks $|p|/\cos\theta$ resolution  &   0.5\\
CC energy calibration   & 0.1 \\
CC $\theta$/$\phi$ resolutions  &  0.2 \\
Background subtraction   & 0.1    \\
$\pi^0$ rejection   & 0.2 \\
Kinematic fit  $\chi^2$ selections  & 0.4 \\

Luminosity    &  0.6    \\

\hline
Sum   &  1.5\\
\hline
\end{tabular}
\end{center}
\label{tbl:sys1}
\end{table}

\section{Results for Cross Sections, $\bm{\sigma(e^+e^-\to\pi^+\pi^-)}$}

The event distributions shown in Fig.~\ref{fig:dist} lead to the cross section for $e^+e^-\to\pi^+\pi^-$ at the energy $\sqrt{s'}=M_{\pi\pi}$, $\sigma_{\pi\pi}(\sqrt{s'})$,  calculated through
\begin{equation} \label{eq409}
\sigma_{\pi\pi}(\sqrt{s'})=\frac{dN_{\pi\pi\gamma_{\mathrm{ISR}}}/d\sqrt{s'}}{(dL^{\mathrm{eff}}_{\mathrm{ISR}}/d\sqrt{s'})(\epsilon(\sqrt{s'}))},
\end{equation}
where $dN_{\pi\pi\gamma_{\mathrm{ISR}}}/d\sqrt{s'}$ is the number of $\pi^+\pi^-\gamma_{\mathrm{ISR}}$ events in the observed $M_{\pi\pi}$ mass spectrum, $dL^{\mathrm{eff}}_{\mathrm{ISR}}/d\sqrt{s'}$ is the effective ISR luminosity, and $\epsilon(\sqrt{s'})$ is the total efficiency to reconstruct these events, determined through Eq.~(\ref{eq:eff}).
The effective ISR luminosity function is given by
\begin{equation} \label{eq:effISR}
\frac{dL^{\mathrm{eff}}_{\mathrm{ISR}}}{d\sqrt{s'}}=L_{ee}\frac{dW}{d\sqrt{s'}}\left(\frac{\alpha(s')}{\alpha(0)}\right)^2,
\end{equation}
where $L_{ee}$ is the $e^+e^-$ luminosity, $dW/d\sqrt{s'}$ is the radiator function calculated up to order $\alpha_{\mathrm{em}}^2$ in Ref.~\cite{w}, and $(\alpha(\sqrt{s'})/\alpha(0))^2$ is the vacuum polarization correction to the fine structure constant~\cite{fine}.

\begin{figure}[!tb]
\begin{center}
\includegraphics[width=3.5in]{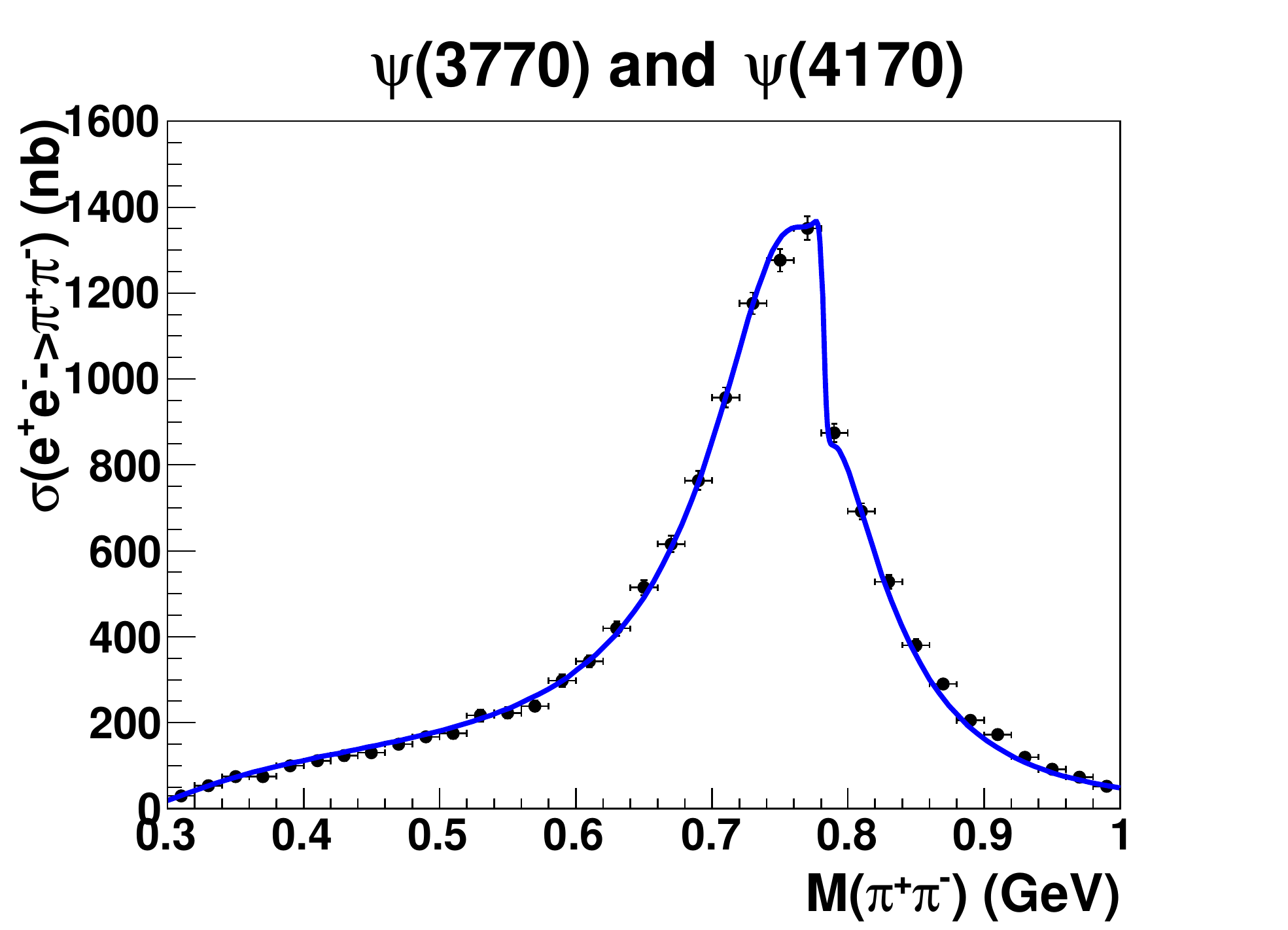} 
\end{center}
\caption[The fit of the measured cross section for $e^+e^-\to\pi^+\pi^-$ for combined data of $\psi(3770)$ and $\psi(4170)$.]
{The fit of the measured cross section for $e^+e^-\to\pi^+\pi^-$ in 20 MeV bin for combined data of $\psi(3770)$ and $\psi(4170)$. 
Statistical errors only.}
\label{fig:fit_cross_section_both}
\end{figure}

In Table~\ref{tbl:cross_section} we list our measured cross sections $\sigma(\pi^+\pi^-)$ as measured separately at $\psi(3770)$ and $\psi(4170)$, and also for the weighted average.
Fig.~\ref{fig:fit_cross_section_both} shows a plot of the average cross sections. 
Both statistical and systematic errors are included in the average cross sections listed in Table~\ref{tbl:cross_section} and plotted in Fig.~\ref{fig:fit_cross_section_both}.

\begin{table*}[!tb]
\caption
{Bare cross section of $e^+e^-\to\pi^+\pi^-$ in 0.02~GeV intervals for data $\psi(3770)$, $\psi(4170)$ and the combination of the two data sets.  
Only statistical errors are shown for data $\psi(3770)$ and $\psi(4170)$. For the weighted average cross section of the two data sets, the first error is statistical and the second is systematic.}
\begin{center}
\begin{tabular}{c|c|c|c}
\hline
$M_{\pi\pi}$ (GeV) &  \multicolumn{3}{c}{$\sigma_{\pi\pi}$ (nb)}    \\
\hline
                   & $\psi(3770)$  & $\psi(4170)$  & wtd average  \\
\hline
0.30--0.32 & $33.6\pm7.8$ & $23.6\pm9.5$ & $29.6\pm6.0\pm0.4$ \\ 
0.32--0.34 & $49.1\pm9.4$ & $64.9\pm14.0$ & $54.0\pm7.8\pm0.8$ \\ 
0.34--0.36 & $71.2\pm11.2$ & $79.0\pm13.8$ & $74.3\pm8.7\pm1.1$ \\
0.36--0.38 & $72.8\pm10.9$ & $80.4\pm14.0$ & $75.7\pm8.6\pm1.1$ \\
0.38--0.40 & $95.8\pm12.3$ & $106.8\pm16.5$ & $99.7\pm9.9\pm1.5$ \\
0.40--0.42 & $110.5\pm13.5$ & $112.9\pm17.7$ & $111.4\pm10.7\pm1.7$ \\
0.42--0.44 & $116.7\pm16.0$ & $133.0\pm19.1$ & $123.4\pm12.3\pm1.9$ \\
0.44--0.46 & $135.1\pm14.8$ & $129.7\pm17.6$ & $132.9\pm11.3\pm2.0$ \\
0.46--0.48 & $157.7\pm15.6$ & $141.2\pm18.6$ & $150.9\pm12.0\pm2.3$ \\
0.48--0.50 & $163.5\pm15.4$ & $175.9\pm21.1$ & $167.8\pm12.4\pm2.5$ \\
0.50--0.52 & $187.3\pm15.9$ & $157.3\pm20.3$ & $176.0\pm12.5\pm2.6$ \\
0.52--0.54 & $219.2\pm18.5$ & $216.9\pm21.7$ & $218.2\pm14.1\pm3.3$ \\
0.54--0.56 & $222.3\pm16.1$ & $228.2\pm21.4$ & $224.5\pm12.9\pm3.4$ \\
0.56--0.58 & $227.3\pm15.1$ & $266.1\pm22.5$ & $239.4\pm12.6\pm3.6$ \\
0.58--0.60 & $289.6\pm17.7$ & $313.3\pm24.3$ & $297.8\pm14.3\pm4.5$ \\
0.60--0.62 & $328.4\pm18.6$ & $369.3\pm24.9$ & $343.0\pm14.9\pm5.1$ \\
0.62--0.64 & $406.9\pm20.3$ & $446.9\pm28.0$ & $420.7\pm16.5\pm6.3$ \\
0.64--0.66 & $506.9\pm22.2$ & $529.1\pm30.8$ & $514.5\pm18.0\pm7.7$ \\
0.66--0.68 & $642.3\pm24.8$ & $579.5\pm30.3$ & $617.1\pm19.2\pm9.3$ \\
0.68--0.70 & $767.9\pm26.8$ & $756.7\pm36.5$ & $764.0\pm21.9\pm11.5$ \\
0.70--0.72 & $941.5\pm29.3$ & $983.0\pm38.3$ & $956.8\pm23.3\pm14.4$ \\
0.72--0.74 & $1194.3\pm32.5$ & $1146.5\pm41.2$ & $1176.0\pm25.5\pm17.6$ \\
0.74--0.76 & $1280.4\pm33.2$ & $1268.6\pm43.3$ & $1276.0\pm26.3\pm19.1$ \\
0.76--0.78 & $1352.6\pm33.6$ & $1347.3\pm45.8$ & $1350.7\pm27.1\pm20.3$ \\
0.78--0.80 & $855.8\pm25.7$ & $911.0\pm36.1$ & $874.4\pm20.9\pm13.1$ \\
0.80--0.82 & $686.0\pm23.0$ & $704.1\pm31.4$ & $692.3\pm18.5\pm10.4$ \\
0.82--0.84 & $545.3\pm20.4$ & $500.7\pm26.4$ & $528.6\pm16.2\pm7.9$ \\
0.84--0.86 & $370.3\pm16.7$ & $400.8\pm23.6$ & $380.5\pm13.6\pm5.7$ \\
0.86--0.88 & $303.6\pm14.8$ & $265.9\pm19.2$ & $289.5\pm11.7\pm4.3$ \\
0.88--0.90 & $220.3\pm12.5$ & $183.6\pm15.6$ & $206.0\pm9.8\pm3.1$ \\
0.90--0.92 & $175.8\pm11.1$ & $166.9\pm14.9$ & $172.6\pm8.9\pm2.6$ \\
0.92--0.94 & $117.1\pm8.9$ & $126.8\pm13.0$ & $120.2\pm7.4\pm1.8$ \\
0.94--0.96 & $91.4\pm7.8$ & $94.5\pm11.2$ & $92.4\pm6.4\pm1.4$ \\
0.96--0.98 & $71.5\pm7.0$ & $80.6\pm10.0$ & $74.4\pm5.7\pm1.1$ \\
0.98--1.00 & $47.3\pm5.5$ & $63.2\pm8.8$ & $51.8\pm4.7\pm0.8$ \\

\hline
\end{tabular}
\end{center}
\label{tbl:cross_section}
\end{table*}

\begin{figure}[!tb]
\begin{center}
\includegraphics[width=3.5in]{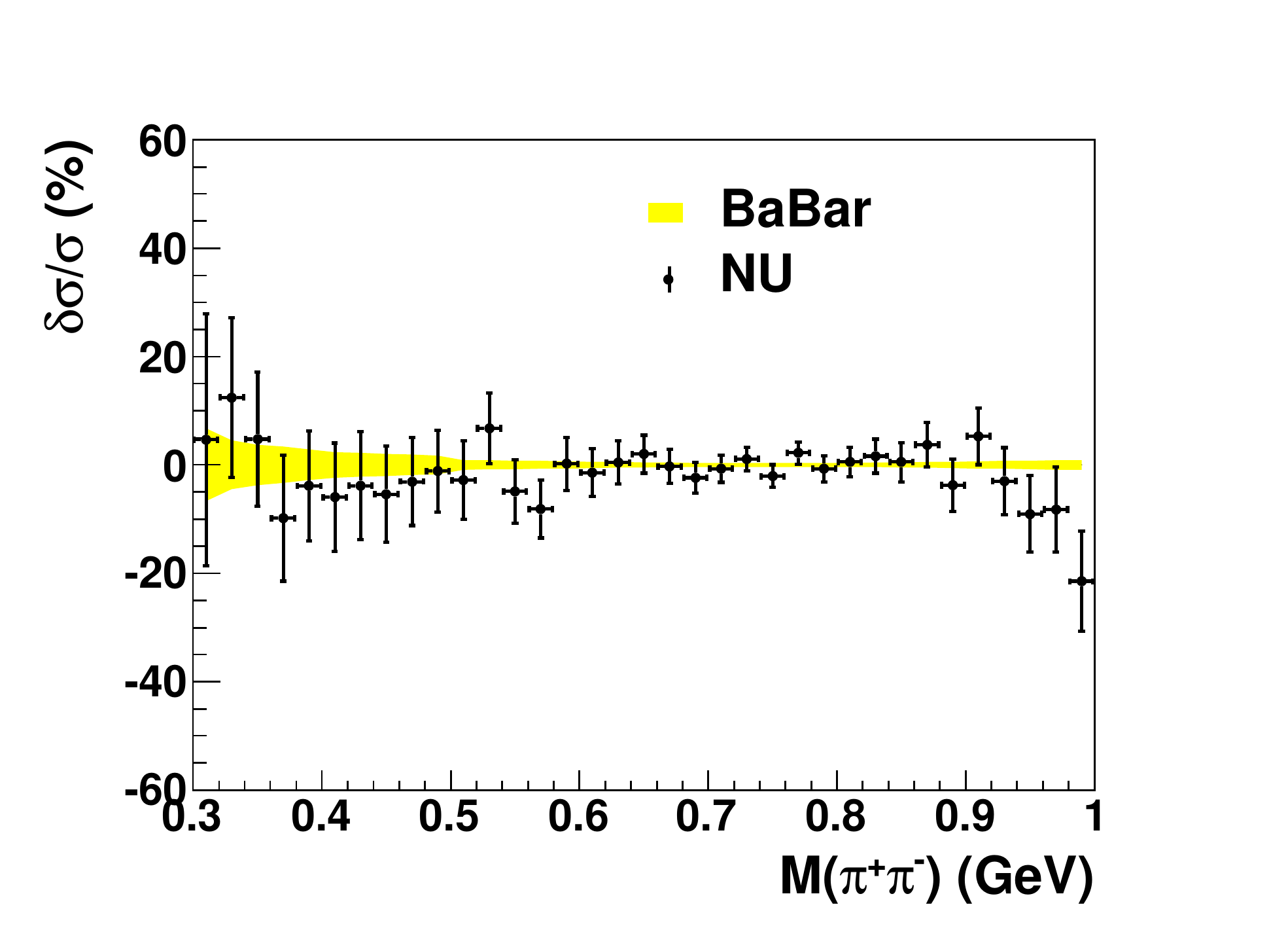}
\includegraphics[width=3.5in]{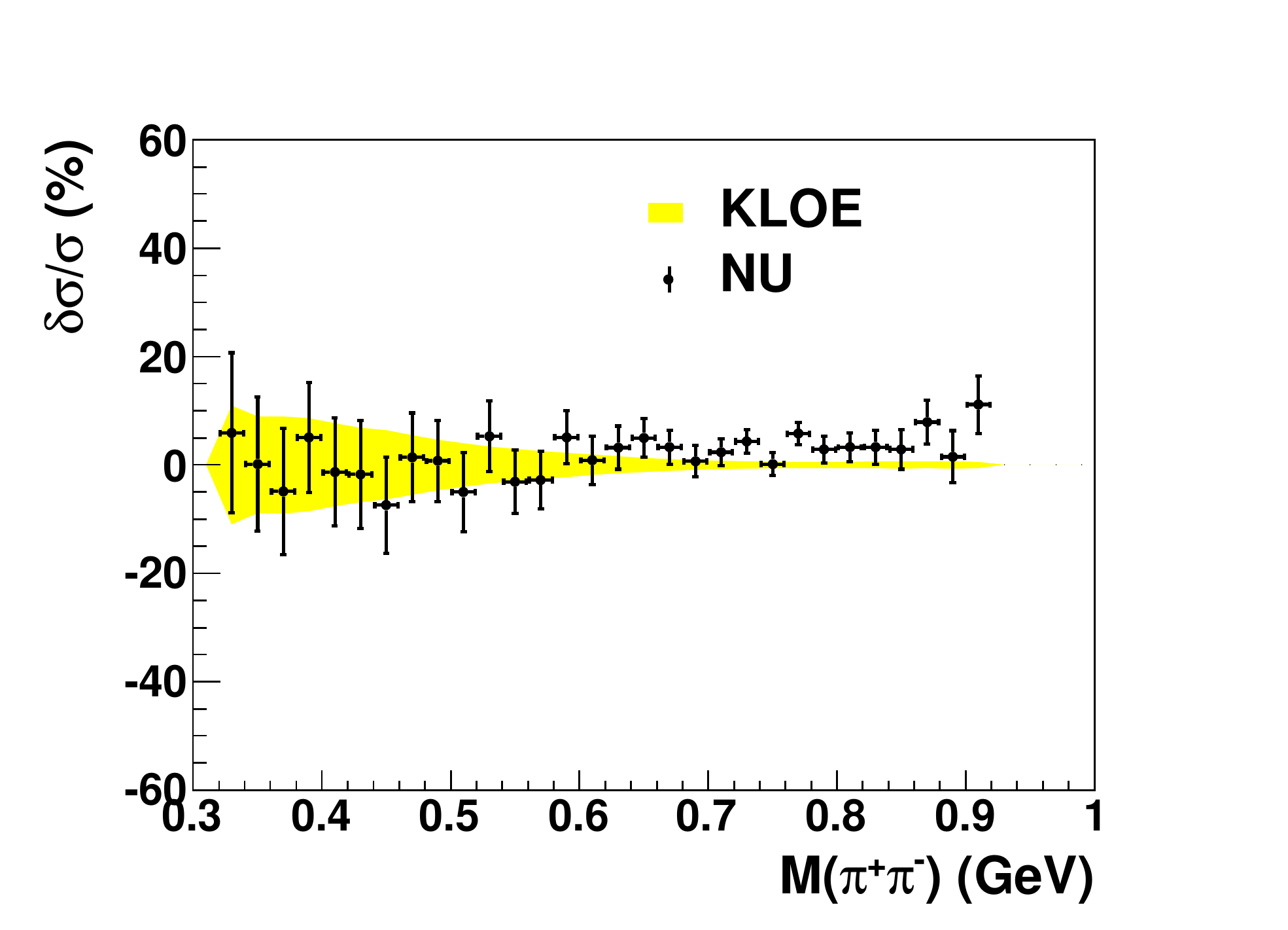}
\end{center}
\caption
{Fractional difference, $\delta(\sigma)/\sigma (\%)$ between our (NU) results and those of BaBar~\cite{babar2} (top) and KLOE~\cite{kloe10} (bottom).
The statistical and systematic uncertainties have been combined in quadrature. The yellow bands show the bounds of BaBar and KLOE uncertainties.
The systematic uncertainties in the NU measurements are discussed in Sec.~\ref{sys}.}
\label{fig:comparison_cross_section_ratio}
\end{figure}

In Fig.~\ref{fig:comparison_cross_section_ratio} we plot the differences between our measured cross sections $\sigma(\mathrm{NU})\pm\delta\sigma$ as percentage differences from the cross sections measured by 
BaBar~\cite{babar2} [left] and KLOE~\cite{kloe10} (right).
The percentage errors in the cross sections measured by BaBar and KLOE are indicated by shaded error bands for each.
In the $\rho$ resonance region where $\sigma(\pi^+\pi^-)$ make the largest contribution to $a_\mu^{\pi\pi,\mathrm{LO}}$ the differences are small in both cases.

As Fig.~\ref{fig:fit_cross_section_both} shows that the cross section is dominated by the well known vector hadron resonances, $\rho(770)$ and $\omega(782)$.
Although it is not the object of this measurement, we can try to fit our $e^+e^-\to\pi^+\pi^-$ cross sections and derive the parameters of these resonances. 
We do so in the framework of the so-called Gounaris-Sakurai (GS) formalism~\cite{gs}, but since our data are confined to $M_{\pi\pi}<1.0$~GeV, we take account of only the first $\rho$-resonance,
$\rho(770)$.
The fit results are listed in Table~\ref{tbl:cross},
along with results from other measurements. The masses of the $\rho$ resonance obtained from our data are in agreement with the values obtained in other experiments and the PDG. The width of $\rho$ is somewhat larger. The $\rho$--$\omega$ interference phase $\phi_\omega$ differs from that obtained by BaBar using higher $\rho$ resonances~\cite{babar2}.

\begin{table*}[!tb]
\caption[The parameters for fitting $e^+e^-\to\pi^+\pi^-$ cross section and comparison to those from other experiments and PDG.]
{The parameters for fitting $e^+e^-\to\pi^+\pi^-$ cross section and comparison to those from other experiments and PDG. Statistical errors only.}
\begin{center}
\begin{tabular}{lcccc}
\toprule
data  & $M_\rho$~(MeV)  & $\Gamma_\rho$~(MeV) & $|c_\omega|\times 10^3$ & $\phi_\omega$~(rad)  \\
\hline
NU &$774.9\pm0.4$  & $154.2\pm0.8$  & $1.8\pm0.2$  & $0.21\pm0.11$   \\
BaBar~\cite{babar2} &$775.02\pm0.31$  & $149.59\pm0.67$  & $1.644\pm0.061$  & $-0.011\pm0.037$   \\
BESIII~\cite{besiii}&$776.0\pm0.4$  & $151.7\pm0.7$  & $1.7\pm0.2$  & $0.04\pm0.13$   \\
CMD-2~\cite{cmd2}&$775.97\pm0.84$  & $145.98\pm0.90$  & --  & $0.182\pm0.067$   \\
SND~\cite{snd} &$774.6\pm0.6$  & $146.1\pm1.7$  & --  & $1.984\pm0.042$   \\
PDG~\cite{pdg2014} &$775.25\pm0.25$  & $149.1\pm0.8$  & --  & --   \\
\hline
\end{tabular}
\end{center}
\label{tbl:cross}
\end{table*}

\section{Results for  Contribution $\bm{a_\mu^{\pi\pi,\mathrm{LO}}}$ to Muon Magnetic Moment Anomaly }

Our measured cross sections $\sigma(e^+e^-\to\pi^+\pi^-)$ lead to the determination of the low energy contribution to the muon magnetic moment anomaly using the dispersion relation in Eq.~(\ref{eq:eq165}).
Our results are:
\begin{equation}\begin{split}
a_\mu^{\pi\pi,\mathrm{LO}}\times 10^{10}&=498.6\pm4.5, ~\mathrm{for}~\psi(3770),\\
                                                                                        &=503.6\pm5.9, ~\mathrm{for}~\psi(4170),\\
                                                                                        &=500.4\pm3.6, ~\mathrm{for~weighted~average}.
\end{split}
\end{equation}
Including systematic uncertainties, our final result is 
\begin{equation}\begin{split}
a_\mu^{\pi\pi,\mathrm{LO}}\times 10^{10}&=500.4\pm3.6~(\mathrm{stat})\pm7.5~(\mathrm{syst})\\
                                                                                        &=500.4\pm8.3~(\mathrm{in~quadrature}).
\end{split}
\end{equation}
Our total uncertainty amounts to $\pm1.7\%$.

\section{Summary and Conclusions}

In summary, we have made an independent precision measurement of the low energy contribution to the muon magnetic moment anomaly $a_\mu\equiv(g-2)/2$ using the initial state radiation method with the $e^+e^-\to\pi^+\pi^-$ data taken at CLEO at the $\psi(3770)$ and $\psi(4170)$ resonances. 
For the range $0.30<M_{\pi\pi}<1.00$~GeV we measure $a_\mu^{\pi\pi,\mathrm{LO}}\times 10^{10}=500.4\pm8.3$.
In Table~\ref{tbl:amu_comp} we compare results obtained by other measurements in this range.
All results are seen to be in agreement within their stated errors, with the largest difference being in the KLOE result.

\begin{table}[!tb]
\caption
{Evaluation of $a_\mu^{\pi\pi,\mathrm{LO}}$ in the range of $0.30<M_{\pi\pi}<1.00$~GeV from different experiments. The errors are from both statistical and systematic sources. 
The results from BaBar, KLOE, CMD-2 and SND are adjusted by BaBar~\cite{babar2}. }

\begin{center}
\begin{tabular}{lcc}
\hline
Experiment & $a_\mu^{\pi\pi,\mathrm{LO}}(\times10^{-10})$ & $M_{\pi\pi}$ range (GeV)\\ 
\hline
NU  &   $500.4\pm8.3$  &0.30--1.00 \\
BaBar      &   $503.6\pm3.4$ &0.30--1.00 \\
KLOE       &   $492.6\pm6.9$  &0.30--1.00\\
CMD-2      &   $496.1\pm3.5$  &0.30--1.00\\
SND        &   $494.6\pm6.5$  &0.30--1.00\\
\hline
Weighted Average   &  $498.7\pm2.1$ &0.30--1.00\\
\hline
BESIII  &   $368.2\pm4.1$  &0.60--0.90 \\
\hline
\end{tabular}
\end{center}
\label{tbl:amu_comp}
\end{table}

As mentioned in the introduction (Sec.~\ref{intro}) the most extensive measurements of $a_\mu^{\pi\pi,\mathrm{LO}}$ have been made by BaBar~\cite{babar2}. They reported:\\
$\sqrt{s}=0.28-1.80~(\mathrm{GeV}):~a_\mu^{\pi\pi,\mathrm{LO}}\times10^{10}=514.09\pm3.82,$ \\
$\sqrt{s}=0.30-1.00~(\mathrm{GeV}):~a_\mu^{\pi\pi,\mathrm{LO}}\times10^{10}=503.56\pm3.38.$\\
The average value of $a_\mu^{\pi\pi,\mathrm{LO}}$ in the region $\sqrt{s}=0.30-1.00~(\mathrm{GeV})$ which we have summarized in Table~\ref{tbl:amu_comp} is\\
$\sqrt{s}=0.30-1.00~(\mathrm{GeV}):~a_\mu^{\pi\pi,\mathrm{LO}}\times10^{10}=498.7\pm2.1.$\\
This is smaller than BaBar's value and has smaller error. 
Since the $\sqrt{s}=0.30-1.00$~GeV region makes the largest contribution to $a_\mu^{\pi\pi,\mathrm{LO}}$ it is interesting to examine the effect of replacing the BaBar value in this region by the average.
This results into $a_\mu^{\pi\pi,\mathrm{LO}}\times10^{10}=509.2\pm2.3$ for $\sqrt{s}=0.28-1.80$~GeV.
This increases the difference between the SM value of $a_\mu\times10^{10}=11659176.4\pm4.9$ and the experimental value of $a_\mu\times10^{10}=11659208.0\pm6.3$~\cite{bnl} to $\Delta a_\mu\times10^{10}=31.6\pm8.0~(4.0\sigma)$, 
and it also increases the tension between $a_\mu^{\pi\pi,\mathrm{LO}}$ and $a_\mu^{\mathrm{had, LO}}=515.2\pm3.6$ from $\tau-$decay, as summarized by BaBar~\cite{babar2}.

\section{Acknowledgement}
 This investigation was done using CLEO data, and as members of the former CLEO Collaboration we thank it for this privilege. This research was supported by the U.S. Department of Energy, Office of Science, Office of Nuclear Physics, under contract DE-FG02-87ER40344.

\end{document}